\documentclass[preprint]{revtex4-1}
\usepackage{epsfig}
\usepackage{graphics}
\usepackage{latexsym}
\usepackage{amsmath}
\usepackage{amssymb}
\usepackage{rotating}
\usepackage{subfigure}
\usepackage{bm}
\usepackage{color}
\usepackage{blindtext}
\usepackage{hyperref}
\usepackage{graphicx}
\usepackage{hhline,multirow,tabularx}  
\begin{document}

\title{Implementation of target mass corrections  and higher-twist  effects in the xFitter framework}

\author{Muhammad Goharipour$^{a}$}
\email{muhammad.goharipour@ipm.ir}

\author{S.~Rostami$^{b}$}
\email{asalrostami.phy@gmail.com}

\affiliation{
$^{a}$School of Particles and Accelerators, Institute for Research in Fundamental Sciences (IPM), P.O. Box 19568-36681, Tehran, Iran\\
$^{b}$Department of Physics, Shahid Rajaee Teacher Training University, Lavizan, Tehran 16788, Iran}

\date{\today}

\begin{abstract}

Knowledge of parton distribution functions (PDFs) at large momentum fraction $ x $ is not only important to study the flavor and spin dynamics of quarks in the nucleon, but also to search for signals of new physics at collider experiments. It is well known now that the nonperturbative QCD effects generally play a more significant role at such kinematic regions. In this work, we present an open-source QCD analysis of PDFs considering target mass corrections (TMCs) and higher-twist (HT) effects which are necessary when the analysis includes also data points from deep inelastic scattering (DIS) with larger values of $ x $ and smaller values of photon virtuality $ Q^2 $. To this aim, we use the xFitter package, as a comprehensive and powerful tool for studying PDFs. We report and discuss our recent activities on the implementation of TMCs and HT effects in the xFitter framework in a user-friendly way, so that the user can switch on or off them, and also choose between different options. We check the validity of the modifications by performing sample analyses of PDFs considering TMCs and HT effects and using a wide range of DIS data.

\end{abstract}


%
\maketitle

\section{Introduction}\label{sec:one} 

Thanks to the high-energy scattering data from various accelerator facilities worldwide, significant advances have been made toward the internal structure of the nucleon using the framework provided by the quantum chromodynamics (QCD). To analyze this huge amount of data, concerted efforts are being made to improve information about the parton distribution functions (PDFs) of the proton~\cite{Butterworth:2015oua,Jimenez-Delgado:2014twa,Dulat:2015mca,Hou:2019efy,Harland-Lang:2014zoa,Ball:2017nwa,Alekhin:2017kpj,Alekhin:2018pai,Abramowicz:2015mha,Rojo:2019uip,Gao:2017yyd,Lin:2017snn,Mottaghizadeha:2017kiw,Sufian:2018cpj,Xu:2019xhk}. 
Various groups are actively involved in extracting PDFs from experimental data, for instance JR~\cite{Jimenez-Delgado:2014twa}, CTEQ~\cite{Dulat:2015mca,Hou:2019efy}, MMHT~\cite{Harland-Lang:2014zoa}, NNPDF~\cite{Ball:2017nwa} and ABMP~\cite{Alekhin:2017kpj,Alekhin:2018pai}.
Although a wide range of experimental data is included in such analyses, the data with low $ W^2 $ (final state invariant mass) and  $ Q^2 $ (hard scale) which are sensitive to the nonperturbative QCD effects are usually cut out, losing some sensitivity at high momentum fractions $ x $.
In return, as an example, one can refer to the analysis of the CTEQ-Jefferson
Lab (CJ) collaboration~\cite{Accardi:2016qay} which has turned its attention to performing a global analysis of PDFs  with the aim of maximally utilizing data from deep inelastic scattering (DIS) at the highest-$ x $ values.
It is worth noting in this context that for extracting PDFs in the large-$ x $ region, it is crucial 
to consider all sources of corrections which may contribute to a comparable magnitude in the theoretical calculations, such as the target mass corrections (TMCs) and higher-twist
(HT) effects~\cite{Accardi:2016qay,Blumlein:2012bf,Accardi:2009br,Brady:2011uy,Khanpour:2017cha}. To be more precise, these supplementary corrections become increasingly important as the hard scale $ Q^2 $ is lowered and $ x $ tends to 1.

Knowledge of PDFs at large $ x $ is important for several reasons. For example, accurate knowledge of the $ u $ and $ d $ quark distributions at high $ x $ is very important at collider experiments in searches for signals of new physics at high $ Q^2 $ values. On the other hand, the flavor and spin dynamics of quarks in the nucleon can be explored by studying the large-$ x $ region, considering this fact that the $ d/u $ ratio at large $ x $ is very sensitive to different mechanisms of spin-flavor symmetry breaking~\cite{Melnitchouk:1995fc,Holt:2010vj,Salajegheh:2017iqp}. It should also be noted that the accurate recognition and calculation of the QCD backgrounds in new physics searches at hadron colliders would not be possible without precise determination of PDFs at large $ x $. 

Motivated by this need for a reliable and consistent
determination of PDFs and their uncertainties using data at the highest-$ x $ values, we
present in this work an open-source QCD analysis of PDFs considering TMCs and HT effects. In fact, the main goal of the present work is providing a user-friendly facility to perform global analysis of PDFs including TMCs and HT effects. To this aim, we use the xFitter package~\cite{Alekhin:2014irh} and modify it to include related theoretical calculations and perform needed jobs correctly. Actually, implementation of TMCs and HT effects in the xFitter package, as a comprehensive and powerful tool for studying PDFs, can be of interest and also useful for further studies in this subject.

The contents of the present paper are as follows. In Sec.~\ref{sec:two}, we discuss the theoretical
framework which is used for calculating TMCs and HT effects on the DIS structure functions. New developments implemented in the xFitter package and different options that user can choose are also explained. This section ends by introducing the fit framework we use in the present work as an example of PDFs analysis considering TMCs and HT effects utilizing the developed xFitter. In Sec.~\ref{sec:three}, we briefly introduce the experimental data included in our sample analyses. Section~\ref{sec:four} is devoted
to present and compare the results obtained from different analyses without and with considering TMCs and HT effects. In Sec.~\ref{sec:five}, some
cross-checks are presented in order to evaluate the validity of the calculations. Finally, we summarize our results and conclusions in Sec.~\ref{sec:six}.
%
\section{THEORETICAL FRAMEWORK}\label{sec:two}
It is well known now that considering TMCs and HT effects in any global analysis of PDFs including DIS data is an essential ingredient, if the analysis contains data points with larger values of $ x $ and smaller values of photon virtuality $ Q^2 $. As mentioned in the Introduction, in this work, we are going to implement these corrections in the xFitter framework~\cite{Alekhin:2014irh}, so that the user can switch on or off TMCs and HT effects in a user-friendly way, and also choose between different options. In the next three subsections, we introduce the theoretical calculations of TMCs and HT effects which are implemented in xFitter and also our fit framework which is utilized in the present analysis. 

\subsection{Target mass corrections (TMCs)}\label{sec:three_A}
Extracting PDFs, especially in the large-$ x  $ region, requires some corrections arise from imposing exact kinematics on twist-2 matrix elements at finite values of 4-momentum transfer $ Q^2 $. These corrections are related to $ x^2 M^2/Q^2 $, where $ M $ is the target mass. In fact, these target mass corrections are needed to consider the effect of mass for particles in the DIS experiment. To be more precise, in such a situation, the mass $ M $ modifies the scattering kinematics and therefore the DIS structure functions.

The TMCs were first calculated by Nachtmann~\cite{Nachtmann:1973mr} in the unpolarized case. It should also be noted that the same method has been applied to the polarized case in Ref.~\cite{Wandzura:1977ce}.
Accordingly, at finite $ Q^2 $, the effects of the target and quark masses modify the
Bjorken variable $  x $ with the light-front momentum fraction. For massless quarks, the parton
light-front fraction is given by the Nachtmann variable,
\begin{equation}
\xi = \frac{2x}{1 + \sqrt{1 + 4 x^2 M^2/Q^2}}.
\label{eq1}
\end{equation}
It is clear from the above formula that TMCs are vanished when  $M^2/Q^2 \to 0$. In other words, for $ x \ll 1 $ or $ Q^2\rightarrow \infty $, two variables $ x $ and $ \xi $ approach each other.

After Nachtmann, Georgi and Politzer in 1976 evaluated TMCs within
the operator product expansion (OPE) for DIS processes~\cite{Georgi:1976ve}. Their approach allows one to express the electroweak structure functions at finite $ Q^2 $ in terms of their $ M^2/Q^2 \rightarrow 0  $ values. Consequently, the target mass corrected structure functions $ F_2^{\textrm{TMC}} $, $ F_L^{\textrm{TMC}} $ and $ F_3^{\textrm{TMC}} $ are given by~\cite{Schienbein:2007gr}
\begin{eqnarray}
F_2^{\textrm{TMC}}(x,Q^2)
&=& {(1+\rho)^2 \over 4 \rho^3} F_2^{(0)}(\xi,Q^2)\
 +\ {3x(\rho^2-1) \over 2\rho^4}
    \int_\xi^1 du\,
    \left[ 1 + {\rho^2-1 \over 2 x \rho} (u-\xi) \right]
    {F_2^{(0)}(u,Q^2) \over u^2},	\nonumber\\
    F_L^{\textrm{TMC}}(x,Q^2)
&=& {(1+\rho)^2 \over 4 \rho} F_L^{(0)}(\xi,Q^2)\
 +\ {x(\rho^2-1) \over \rho^2}
    \int_\xi^1 du\,
    \left[ 1 + {\rho^2-1 \over 2 x \rho} (u-\xi) \right]
    {F_2^{(0)}(u,Q^2) \over u^2},	\nonumber\\
        F_3^{\textrm{TMC}}(x,Q^2)
&=& {(1+\rho) \over 2 \rho^2} F_3^{(0)}(\xi,Q^2)\
 +\ {(\rho^2-1) \over 2\rho^3}
    \int_\xi^1 du
    {F_3^{(0)}(u,Q^2) \over u},	
\label{eq2}    
\end{eqnarray}
where $F_{2,3,L}^{(0)}$ are the structure functions in the $M^2/Q^2 \to 0$
limit and
\begin{equation}
\rho^2 = 1 + \frac{4 x^2 M^2}{Q^2}.
\label{eq3}
\end{equation}

In addition to the OPE, there is an alternative formulation of TMCs~\cite{Ellis:1982cd,Collins:1989gx} within the collinear factorization framework which implements TMCs directly in momentum space. An important advantage of this approach is that it could also be applied to processes other than inclusive DIS~\cite{Accardi:2009md,Guerrero:2015wha}, though it has been shown that two approaches are equivalent in $ {\cal O}(1/Q^2) $. Overall, various prescriptions have been proposed so far for TMCs, using different approximations to the OPE and collinear factorization methods~\cite{Schienbein:2007gr,Aivazis:1993kh,Kretzer:2002fr,Accardi:2008ne,Steffens:2012jx}. There are also some efforts to implement TMCs in Mellin space through the contour integral representations both for the unpolarized~\cite{Georgi:1976ve} and polarized cases~\cite{Blumlein:1998nv}. The latter also includes the corresponding relations for the twist-3 contributions to the polarized structure functions, though such contributions have also been derived in Ref.~\cite{Piccione:1997zh}. It should be noted that TMCs can also be considered in nonforward~\cite{Belitsky:2001hz,Geyer:2004bx} and diffractive~\cite{Blumlein:2008di} scattering.

Since the evaluation of the convolution integrals in Eq.~(\ref{eq2}) can be time consuming,
it is useful to have an approximate version of the target mass corrected structure functions.
Such approximations have also been presented in Ref.~\cite{Schienbein:2007gr} which are in very good agreement with the exact results. 
For example, to achieve the approximate version of $ F_2^{\textrm{TMC}} $ in Eq.~(\ref{eq2}), the integral terms $\int_{\xi}^{1}du F_{2}^{(0)}(u,Q^{2})/u^{2} $ and $\int_{\xi}^{1}du (u-\xi) F_{2}^{(0)}(u,Q^{2})/u^{2} $ are replaced with $(F_{2}^{(0)}(\xi) / \xi) (1-\xi)$ and $ F_{2}^{(0)}(\xi)(-\ln \xi - 1 +\xi)$, respectively, which can be easily evaluated to obtain an upper bound for the contribution of the nonleading terms. In this way, the target mass corrected structure functions can be written in the following forms (please note that the related equation for $ F_L^{{\rm TMC}} $ has not been presented explicitly in Ref.~\cite{Schienbein:2007gr}):
\begin{eqnarray}
F_{2}^{{\rm TMC}}(x,Q^{2}) &\simeq& \frac{x^{2}}{\xi^{2}\rho^{3}}F_{2}^{(0)}(\xi)
\bigg[1+\frac{6\mu x \xi}{\rho}(1-\xi)^2 \bigg],	\nonumber\\
F_{L}^{{\rm TMC}}(x,Q^{2}) &\simeq& \frac{x^{2}}{\xi^{2}\rho}F_{2}^{(0)}(\xi)
\bigg[\frac{F_{L}^{(0)}}{F_{2}^{(0)}}+\frac{4\mu x \xi}{\rho}(1-\xi)+\frac{8\mu^2 x^2 \xi^2}{\rho^2}(-\ln \xi  -1+\xi) \bigg],	\nonumber\\
F_{3}^{{\rm TMC}}(x,Q^{2}) &\simeq& \frac{x}{\xi \rho^{2}} F_{3}^{(0)}(\xi)
\bigg[1-\frac{\mu x \xi}{\rho} (1-\xi) \ln \xi \bigg]\,,
\label{eq4}
\end{eqnarray}
where $ \mu $ is defined as $ \mu=\frac{M^2}{Q^2} $. From a computational point of view, the approximate formulas of Eq.~(\ref{eq4}) can be calculated very fast, but calculation of the exact relations of Eq.~(\ref{eq2}) can be time consuming because they include also integral terms. The situation gets worse if the heavy flavor contributions $ F_{2,L}^{c,b} $ to the structure functions should be calculated depending on the selected heavy quark scheme. Therefore, using the approximate TMCs can be a good option. 

According to the above explanations, we have added two options for calculating TMCs to the xFitter package: ``\texttt{Approximate}" which uses approximate formulas of Eq.~(\ref{eq4}) and ``\texttt{Exact}" which uses exact relations of Eq.~(\ref{eq2}). However, since a version of TMCs has been implemented in the \texttt{APFEL} package~\cite{Bertone:2013vaa} and, on the other hand, \texttt{APFEL} already exists in xFitter, it is also of interest to add TMCs option on xFitter according to the \texttt{APFEL} calculations. In fact, before the present work, it had not been used in xFitter by default, so that there was no any option for users to switch on or off TMCs. Consequently, at the present, we have modified xFitter to include TMCs in a user-friendly way so that the user can switch between three options: 
\begin{itemize}
\item \texttt{Exact} TMCs according to Eq.~(\ref{eq2})
\item \texttt{Approximate} TMCs according to Eq.~(\ref{eq4})
\item \texttt{APFEL} TMCs using formulas implemented in the \texttt{APFEL} package.
\end{itemize}
However, some points should be noticed. First, the xFitter calculations of the DIS structure functions, and then the reduced cross sections, have been classified according to the flavor number schemes, namely, zero-mass variable flavor number scheme (\texttt{ZM-VFNS})~\cite{Collins:1986mp}, fixed-flavor number scheme (\texttt{FFNS})~\cite{Laenen:1992cc,Laenen:1992zk}, \texttt{ABM FFNS}~\cite{Alekhin:2010sv} \texttt{RT}~\cite{Thorne:1997ga,Thorne:2006qt,Thorne:2012az}, \texttt{(S)ACOT}~\cite{Aivazis:1993kh,Aivazis:1993pi,Kramer:2000hn,Kretzer:2003it} and \texttt{FONLL}~\cite{Cacciari:1998it,Forte:2010ta}. Note that the last three  are a general-mass variable flavor number scheme (GM-VFNS). Since \texttt{APFEL} just works with \texttt{FONLL} scheme, for calculating TMCs using \texttt{APFEL} package, the user should choose \texttt{FONLL} and also \texttt{DGLAP\_APFEL} for theory type (collinear evolution with \texttt{APFEL}) in the ``steering.txt" file of xFitter. Second, the user can choose any heavy flavor scheme if the \texttt{Exact} or \texttt{Approximate} TMCs is selected. Third, although using the \texttt{Exact} option for calculating TMCs is superior to the \texttt{APFEL} option due to above-mentioned reasons, but \texttt{APFEL} has an advantage to perform related calculations up to 3 times faster. Fourth, \texttt{APFEL} uses a relation for $ F_2^{\rm TMC} $ which has a less term than the relation presented in Eq.~(\ref{eq2}). To be more precise, the \texttt{APFEL} procedure can be considered equivalent to the \texttt{Exact} procedure, with a missing term in $ F_2^{\rm TMC} $. Nevertheless, as we checked it, using \texttt{APFEL} and \texttt{Exact} options leads to the results which are in  excellent agreement with each other, though \texttt{Exact} leads to a somewhat better fit of DIS data than \texttt{APFEL}.

\subsection{Higher twist (HT) effects}\label{sec:three_B}
As mentioned in the Introduction, in addition to the TMCs, HT effects must also be taken into account in
any analysis containing experimental data at low $ Q^2 $ and especially at large $ x $.
In the OPE framework, HT corrections to DIS processes are associated with matrix elements of operators involving multiquark or quark and gluon fields~\cite{Jaffe:1982pm,Ellis:1982wd,Shuryak:1981kj,Bukhvostov:1985rn,Balitsky:1987bk}. Although the details of these corrections are not yet fully understood, it is well known now that they are proportional to $ 1/Q^2 $ to the structure functions and become increasingly important as $ Q^2 $ is lowered and $ x $ tends to 1. Because dynamical higher twists involve nonperturbative multiparton interactions, there are theoretical difficulties of controlling power corrections in effective theories, so that it is difficult to quantify the shape of the higher twist terms from first principles. Consequently, HT effects are usually determined phenomenologically from experimental data. To be more precise, for extracting leading twist PDFs one can parametrize the HT contributions by a phenomenological form and fit the unknown parameters to the data~\cite{Accardi:2016qay,Accardi:2009br,Alekhin:2007fh,Pumplin:2002vw}. It is also possible to extract the $ Q^2 $ dependence of HT contributions by fitting a phenomenological form in individual bins in $ x $~\cite{Virchaux:1991jc,Martin:1998np,Blumlein:2008kz,Alekhin:2002fv}. Moreover, the HT corrections can be obtained from the renormalon formalism~\cite{Hadjimichef:2019vbb}. It is worth noting in this context that considering the HT effects in the analysis of the diffractive PDFs can also lead to some improvements in the fit quality and change the shape of the extracted densities at some kinematic regions~\cite{Maktoubian:2019ppi}.

It should be mentioned that xFitter has already an option for considering HT according to GBW dipole model~\cite{GolecBiernat:1998js,GolecBiernat:1999qd}. Although the dipole picture provides a natural description of the QCD reaction in the low $ x $ region, and then presents a good description of HERA data, it is not a suitable case for fixed target data. Consequently, in the present work, we are going also to add another option to xFitter for considering HT effects in a way that is appropriate for fixed target data. In this way, following the procedure used in the CJ15 analysis~\cite{Accardi:2016qay}, we parametrize HT effects as a function of $ x $ and use the commonly used phenomenological form for the total structure function as follows:
\begin{equation}
 F_2(x,Q^2)= F_2^{\textrm{LT}}(x,Q^2) \bigg(1+\frac{C_{\textrm{HT}}(x)}{Q^2}\bigg),
 \label{eq5}
\end{equation}
which includes higher twists but also other residual power corrections.
In the above formula, $ F_2^{\textrm{LT}} $ is the leading twist structure function including TMCs and the function $ C_{\textrm{HT}} $ is parametrized as
\begin{equation}
C(x)= h_0 x^{h_1}(1+h_2x),
\label{eq6}
\end{equation}
where $ h_1 $ controls the rise of $ C_{\textrm{HT}} $ at large $ x $ and 
$ h_2 $ allows for the possibility of negative HT at smaller $ x $.

In the view of code, we have added another type of HT in xFitter, namely ``\texttt{CJ15}", to already available type ``\texttt{Twist4}" which is according to the GBW dipole model (without charm). Therefore, if a user uses the \texttt{CJ15} option for ``\texttt{HiTwistType}" in the ``steering.txt" file, three parameters $ h_0 $, $ h_1 $ and $ h_2 $ should be set as extra \texttt{MINUIT}~\cite{James:1975dr} parameters. It should be noted that the \texttt{twist4} option in xFitter considers the twist-4 contribution to neutral current (NC) reduced cross section, while the \texttt{CJ15} option considers HT on the $ F_2 $ structure function.

\subsection{Fit framework}\label{sec:three_C}
As mentioned before, the main goal of the present work is providing a user-friendly facility to perform global analysis of PDFs including TMCs and HT effects by developing the open-source program xFitter~\cite{Alekhin:2014irh}. In this way, in the present study, we just perform some simple QCD analyses of PDFs at next-to-next-to-leading order (NNLO) using developed xFitter to test the codes and check the results obtained. For sample analyses  performed in the next section, we use the \texttt{HERAPDF} forms~\cite{Abramowicz:2015mha} for the generic parametrization of PDFs at the input scale $ \mu^2_0 = 1.69  $ GeV$ ^2 $. It should be noted that the optimal parametrizations for PDFs can be found through a parametrization scan as described in~\cite{Aaron:2009aa}. In this way, the valence
quarks $ xu_v(x) $ and $ xd_v(x) $, the antiquarks $ x\bar{U}(x)=x\bar{u}(x) $ and $ x\bar{D}(x)=x\bar{d}(x)+x\bar{s}(x) $, and gluon $ xg(x) $ are parametrized as~\cite{Abramowicz:2015mha}
\begin{eqnarray}
xg(x) &=   & A_g x^{B_g} (1-x)^{C_g} - A_g' x^{B_g'} (1-x)^{C_g'}  ,  \nonumber\\
xu_v(x) &=  & A_{u_v} x^{B_{u_v}}  (1-x)^{C_{u_v}}\left(1+E_{u_v}x^2 \right) ,\nonumber\\
xd_v(x) &=  & A_{d_v} x^{B_{d_v}}  (1-x)^{C_{d_v}} , \nonumber\\
x\bar{U}(x) &=  & A_{\bar{U}} x^{B_{\bar{U}}} (1-x)^{C_{\bar{U}}}\left(1+D_{\bar{U}}x\right) , \nonumber\\
x\bar{D}(x) &= & A_{\bar{D}} x^{B_{\bar{D}}} (1-x)^{C_{\bar{D}}}.
\label{eq7}
\end{eqnarray}
As usual, the normalization parameters $A_{u_v}, A_{d_v}$ and $  A_g $ are constrained 
by the well-known quark number and momentum sum rules. $B$ parameters for sea quark distributions, namely $B_{\bar{U}}$ and $B_{\bar{D}}$, are set to be equal, $B_{\bar{U}}=B_{\bar{D}}$.
As well, constraint condition $A_{\bar{U}}=A_{\bar{D}} (1-f_s)$ is applied
to ensure that $x\bar{u} \rightarrow x\bar{d}$ as $x \rightarrow 0$. The strange-quark distribution is taken to be proportional to $ x\bar{D} $ as $x\bar{s}= f_s x\bar{D}$ at $\mu^2_0$ and the value of $f_s$ 
is chosen as $ f_s= 0.4 $~\cite{Martin:2009iq,Nadolsky:2008zw,Aad:2012sb}. It should be also noted that we impose kinematic cut $ Q^2 > 1.69 $ GeV$ ^2 $ on all DIS data and $ W^2 > 3.0 $ GeV$ ^2 $ on all fixed target data, since we are going to study the impact of TMCs and HT effects on extracted PDFs (we study in detail the impact of different choices for $ W^2 $ cut on the extracted PDFs in Sec.~\ref{sec:five}). The strong coupling constant at the Z boson mass scale is taken as $ \alpha_s(M_{\textrm{Z}}^2)= 0.118 $. For including TMCs wherever needed, we use the \texttt{Exact} option (see Sec.~\ref{sec:three_A}). Finally, to take into account the effects of heavy quark masses, we choose the \texttt{FONLL-C} scheme as our analyses are performed at NNLO.

%
\section{Experimental Data}\label{sec:three}
By introducing the theoretical framework in the previous section, in this section, we briefly introduce the experimental data which are used in our sample analyses. As mentioned before, since the main goal of the present work is developing xFitter to include TMCs and HT effects and the study of their impacts on the extracted PDFs, and not performing a comprehensive global analysis of PDFs, we just consider all available DIS data which are sensitive to TMCs and HT effects. In this way, in addition to the combined inclusive~\cite{Abramowicz:2015mha} and heavy-flavor production cross sections~\cite{Abramowicz:1900rp,Abramowicz:2014zub} at HERA and fixed target data from BCDMS collaboration~\cite{Benvenuti:1989rh} which are available in the xFitter package, we have also added and used fixed target data from SLAC~\cite{Whitlow:1991uw}, NMC~\cite{Arneodo:1996qe}, E665~\cite{Adams:1996gu} and Jefferson Lab~\cite{Malace:2009kw}. All experimental datasets used in the present study have been listed in the first column of Table~\ref{tab:chi2}.

%
\section{Results}\label{sec:four}
To study the impact of TMCs and HT effects on the extracted PDFs and quality of the fit, and also check the validity of the calculations, we perform three analyses considering an acceptable range of the DIS data introduced in the previous section. The first analysis we called ``\texttt{Base}" is performed without considering TMCs and HT in a phenomenological framework introduced in Sec.~\ref{sec:three_C}. For the second analysis, namely ``\texttt{TMC}", we include also TMCs with the formulation described in Sec.~\ref{sec:three_A} to study their impact on PDFs and the $ \chi^2 $ value of individual datasets. Here we use  the \texttt{Exact} option which leads to the most accurate result. It should be again noted that if one chooses the \texttt{APFEL} option, the same results are obtained though it leads to a somewhat worse fit of DIS data than \texttt{Exact}. Finally, in the third analysis, namely ``\texttt{TMC\&HT}", we consider also HT effects with formulation described in Sec.~\ref{sec:three_B}. Please note that for all analyses  performed in this section we choose \texttt{FONLL-C} for the heavy quark scheme.
   
Table~\ref{tab:chi2} contains the results of the three aforementioned analyses which have been presented
in the third, fourth and fifth columns, respectively. For each dataset included in the fit, we have presented the value of $ \chi^2 $ divided by number of data points, $\chi^2$/$ N_{\textrm{pts.}} $. The value of total $ \chi^2 $ per number of degrees of freedom, $\chi^2$/d.o.f., has also been presented for each analysis in the last row of the table. Although the results obtained clearly show a decrease in the value of total $\chi^2$/d.o.f. by including TMCs and then HT effects from 1.86 to 1.39 and 1.35, respectively,  there are some points that should be noted. 

First, comparing the results of \texttt{Base} and \texttt{TMC} analyses, one can find that the $ \chi^2 $ value has reduced or remained unaffected for more datasets after considering TMCs. The largest decreases are related to the SLCA, JLab and BCDMS datasets. Second, the only considerable increase occurring in the $ \chi^2 $ value is related to the ``HERA1+2 CCem" date set. In fact, such degradation as well as considerable improvement in $ \chi^2 $ for ``HERA1+2 NCep 920" and ``HERA1+2 NCem" samples are somewhat surprising. Actually, we do not expect that TMCs have a significant impact on HERA data which  belong to different kinematics rather than ones in which TMCs are effective. Since it may be due to a change of PDFs and not because of including TMCs, one of the best ways to check the results and evaluate the validity of TMC calculations is performing a fit to HERA data separately. We investigated it by performing two analyses of HERA datasets solely, one without considering TMCs and the other by turning TMCs on, and found that the results of the two analyses are exactly the same. This fact indicates that TMC calculations which have been implemented in the xFitter work properly, and the considerable improvement in $ \chi^2 $ for ``HERA1+2 NCep 920" and ``HERA1+2 NCem" samples, and also significant degradation for ``HERA1+2 CCem" dataset is really due to change of PDFs. Another way to investigate this issue and check the validity of calculations is making predictions and providing a table of partial $ \chi^2 $ for all datasets included in Table~\ref{tab:chi2} using the  same input PDFs for two situations: with and without considering TMCs. Please note that in this way, there is no need to perform a fit. Overall, by comparing the $ \chi^2 $ values in these two situations one can find those datasets which are sensitive to TMCs. We performed it and found that the $ \chi^2 $ values of HERA datasets do not change by turning TMCs on and they are not sensitive to TMCs as expected. This result also shows that TMC calculations are performed accurately and the reason for considerable improvement and deterioration in the $ \chi^2 $ values of the HERA datasets is really the change of PDFs. 

Another point should be noted is concerning the rather poor fit
quality for the JLab data with a $ \chi^2 $ per point of $\chi^2$/$ N_{\textrm{pts.}}\sim 2 $ (the value of $ \chi^2 $ is 292 for 142 data points) which is larger than the corresponding one obtained, e.g., from the CJ analysis~\cite{Accardi:2016qay}. Actually, compared with other global analyses like CJ15,
we have used parametrizations with less flexibility (especially for the case of sea quark
distributions) considering this fact that we have not aimed to perform a comprehensive
global analysis of PDFs and just wanted to check our TMC and HT implementations in
the xFitter and study the impact of them on the extracted PDFs. So, it is obvious that
using less flexible parametrizations may lead to a relatively larger value of $ \chi^2 $ for each dataset included in the analysis and thus the value of total $ \chi^2 /\mathrm{d.o.f.} $.

As a last point, it should be noted that the impact of HT on the value of total $\chi^2$ is decreased when we turn it on after considering TMCs. In fact, the results obtained clearly show that considering TMCs solely can include the nonperturbative effects to a large extent and then HT helps to improve results even further. We found that such a situation is also established if one first turns on HT and  then includes TMCs, which indicates that considering HT effects solely can also include the nonperturbative effects to a large extent. However, the best option is considering TMCs and HT effects simultaneously.  

Table~\ref{tab:par} makes a comparison between the optimum parameters obtained from three analyses \texttt{Base}, \texttt{TMC} and \texttt{TMC\&HT} which have been presented in the second, third and fourth columns, respectively. Note that the parameters in bold have been fixed in the fit. As can be seen, including TMCs and HT effects leads to a significant change in the value of optimum parameters for some flavors and at some kinematic region in momentum fraction $ x $. This can lead to some changes in the shape and behavior of the extracted PDFs. Another point that should be noted is that the HT parameters $ h_0 $, $ h_1 $ and $ h_2 $ have been well constrained by the related data points included in the analysis.

Figure~\ref{fig:fig1} shows a comparison between three aforementioned analyses \texttt{Base} (solid curve), \texttt{TMC} (dashed curve) and \texttt{TMC\&HT} (dotted curve) where we have plotted the $ u_v(x) $ and $ d_v(x) $ valence, gluon $ g(x) $, sum of sea quark $ \Sigma(x) $ and the ratio of $ d_v(x)/u_v(x) $ distributions at the initial scale $ Q^2_0=1.69 $ GeV$ ^2 $ with their uncertainties. As can be seen, the $ u_v(x) $ and $ d_v(x) $ valence distributions have been most affected by including TMCs and HT as expected, since these distributions are dominant at larger values of $ x $ than the others where TMCs and HT effects are also effective. Overall, by including TMCs and HT effects, the $ u_v(x) $ and $ d_v(x) $ distributions are increased and decreased, respectively, while the gluon and sea quark distributions do not change significantly. The impact of TMCs and HT on valence distributions is better characterized in the last panel where we have compared the ratio of $ d_v(x) $ to $ u_v(x) $ for three analyses. One can clearly see an enhancement and reduction at medium and small $ x $ regions, respectively, which can be attributed to the inclusion of the TMCs and HT for the first one and quark number sum rules for the second one.

To investigate the impact of TMCs and HT on the extracted PDFs at higher values of $ Q^2 $, as an example, in Fig.~\ref{fig:fig2} we have shown the ratio of $ u_v(x) $, $ d_v(x) $, $ \Sigma(x) $ and $ d_v(x)/u_v(x) $ distributions obtained from three analyses \texttt{Base}, \texttt{TMC} and \texttt{TMC\&HT} at $ Q^2=10 $ GeV$ ^2 $ to those obtained from the \texttt{Base} analysis as a reference fit, so that the bands represent the related uncertainties. Actually, we have presented here these results as ratio plots to study the impact of TMCs and HT effects with more details. As can be seen, considering TMCs and HT in the analysis of PDFs modifies the valence distributions more than other flavors almost at all values of $ x $. Although the gluon distribution has some changes at medium to high $ x $ regions, the sum of sea quark distribution has only affected at very large $ x $ values. There is no  significant change for both of them at small $ x $ values. 

Generally, the results presented in this section clearly indicate the importance of TMCs and HT effects in the analysis of DIS data when it includes also data points with a larger value of $ x $ and  a smaller value of $ Q^2 $. Actually, considering such corrections can significantly change the extracted PDFs, especially the valence distributions, and improve the agreement between the theory and experiment. Consequently, implementing TMCs and HT effects in the xFitter package, as a comprehensive and powerful tool for studying PDFs, is of interest and also useful for further studies in this subject. The developed version of xFitter which includes TMCs and HT effects with formulation and instruction introduced in the present paper is public and openly accessible on the \texttt{Gitlab} repository~\cite{gitlab}.

%
\section{cross-checks}\label{sec:five}
\begin{figure}[t!]
\centering
\includegraphics[width=0.47\textwidth]{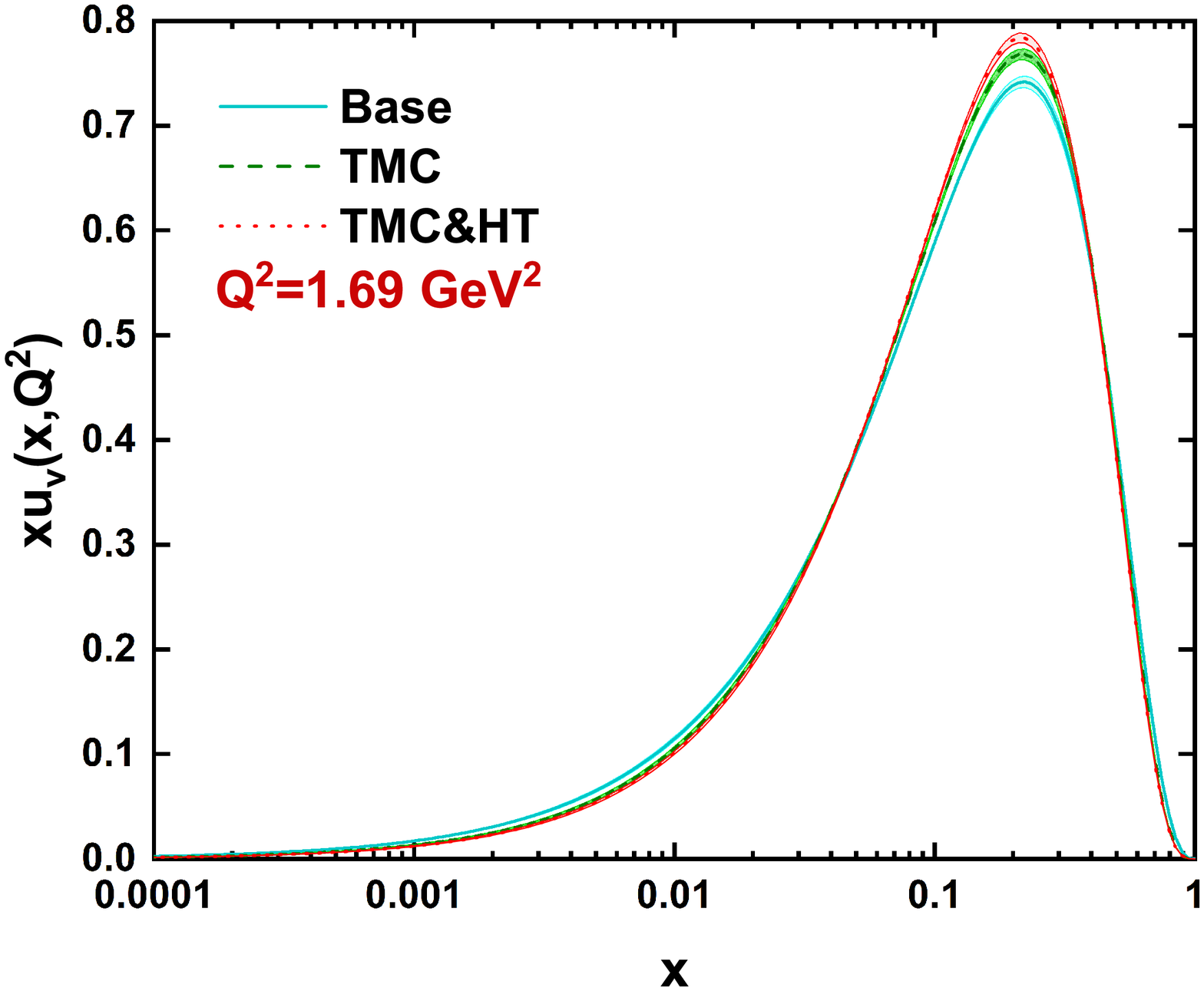}
\includegraphics[width=0.47\textwidth]{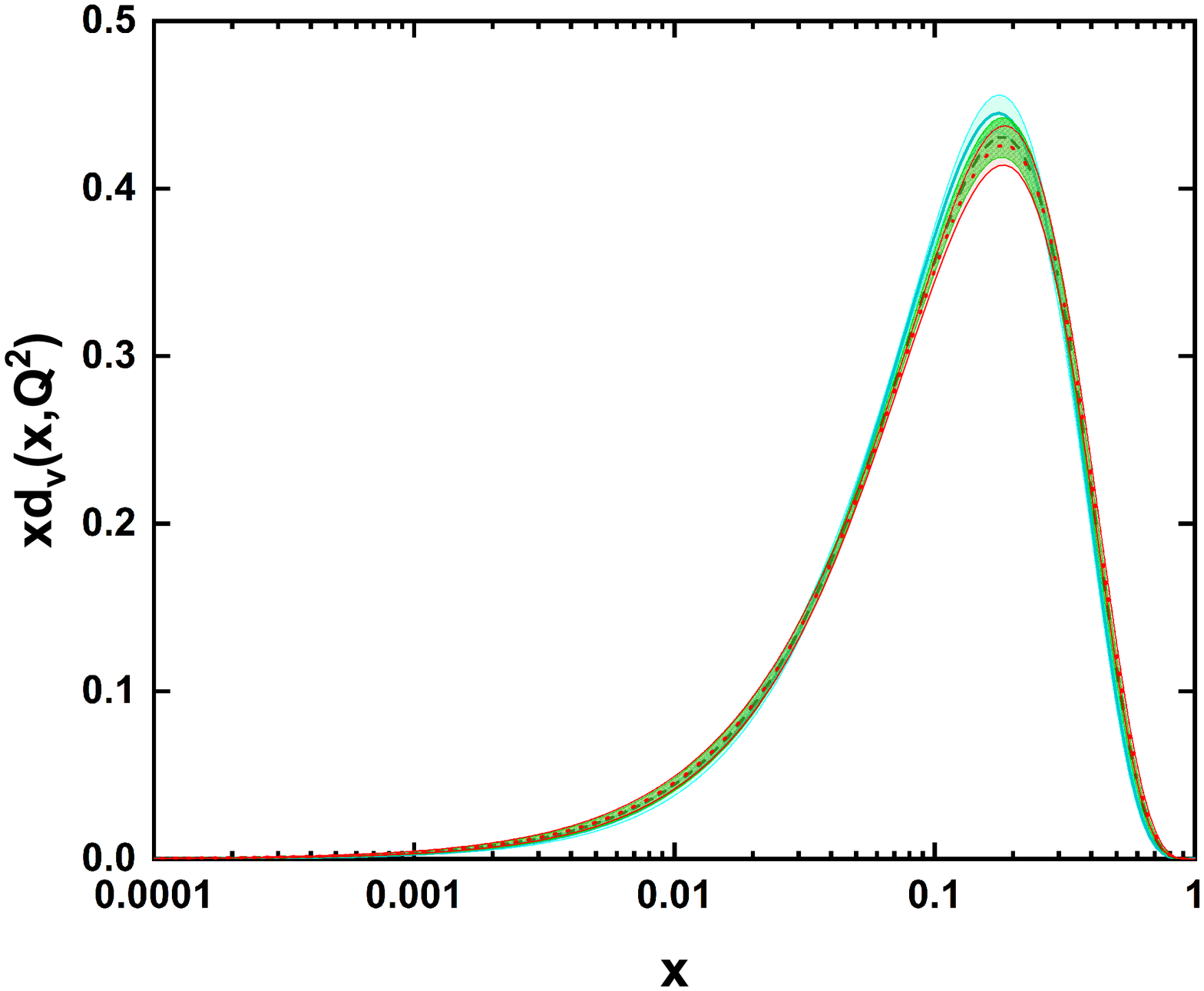}
\includegraphics[width=0.47\textwidth]{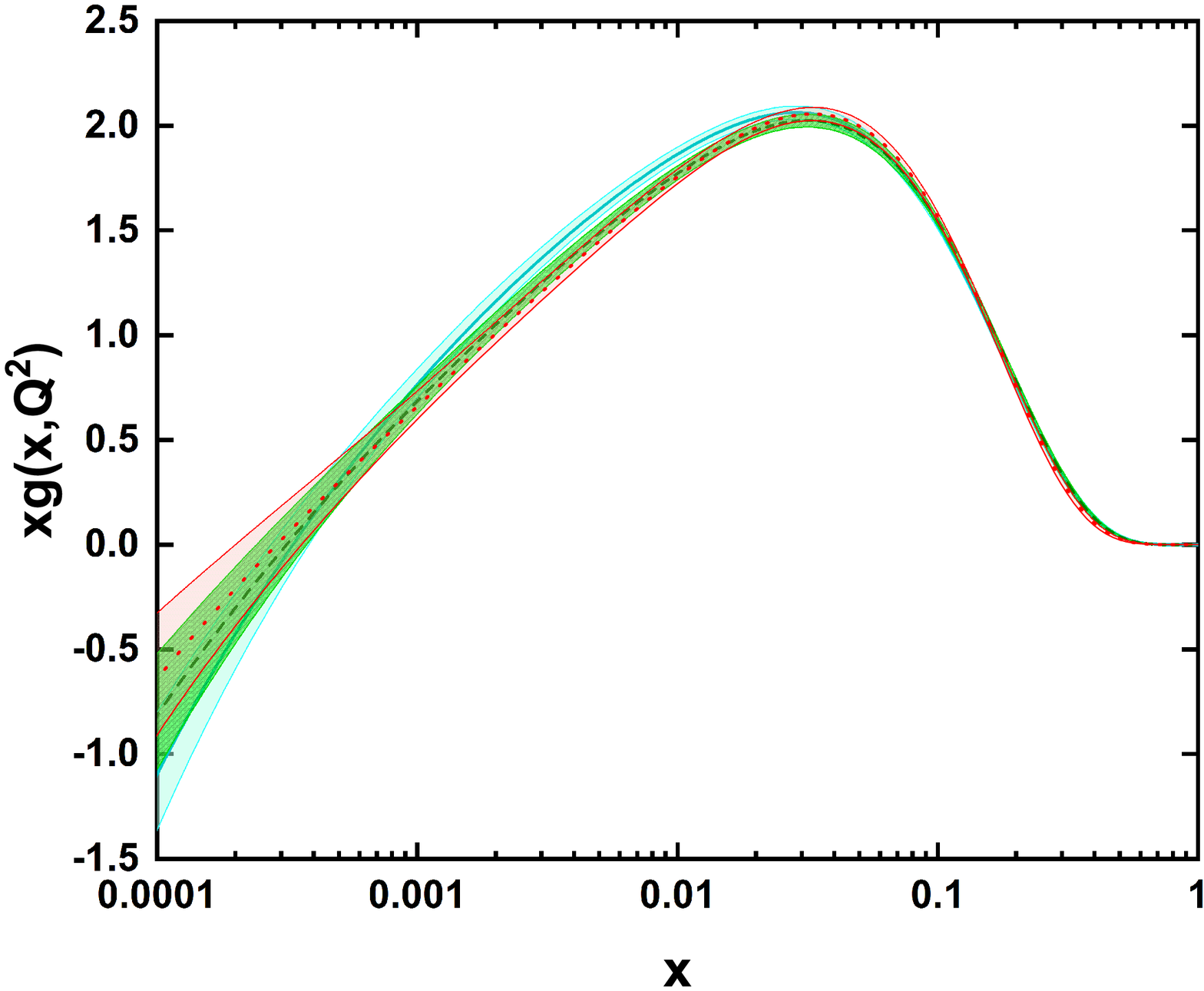}
\includegraphics[width=0.47\textwidth]{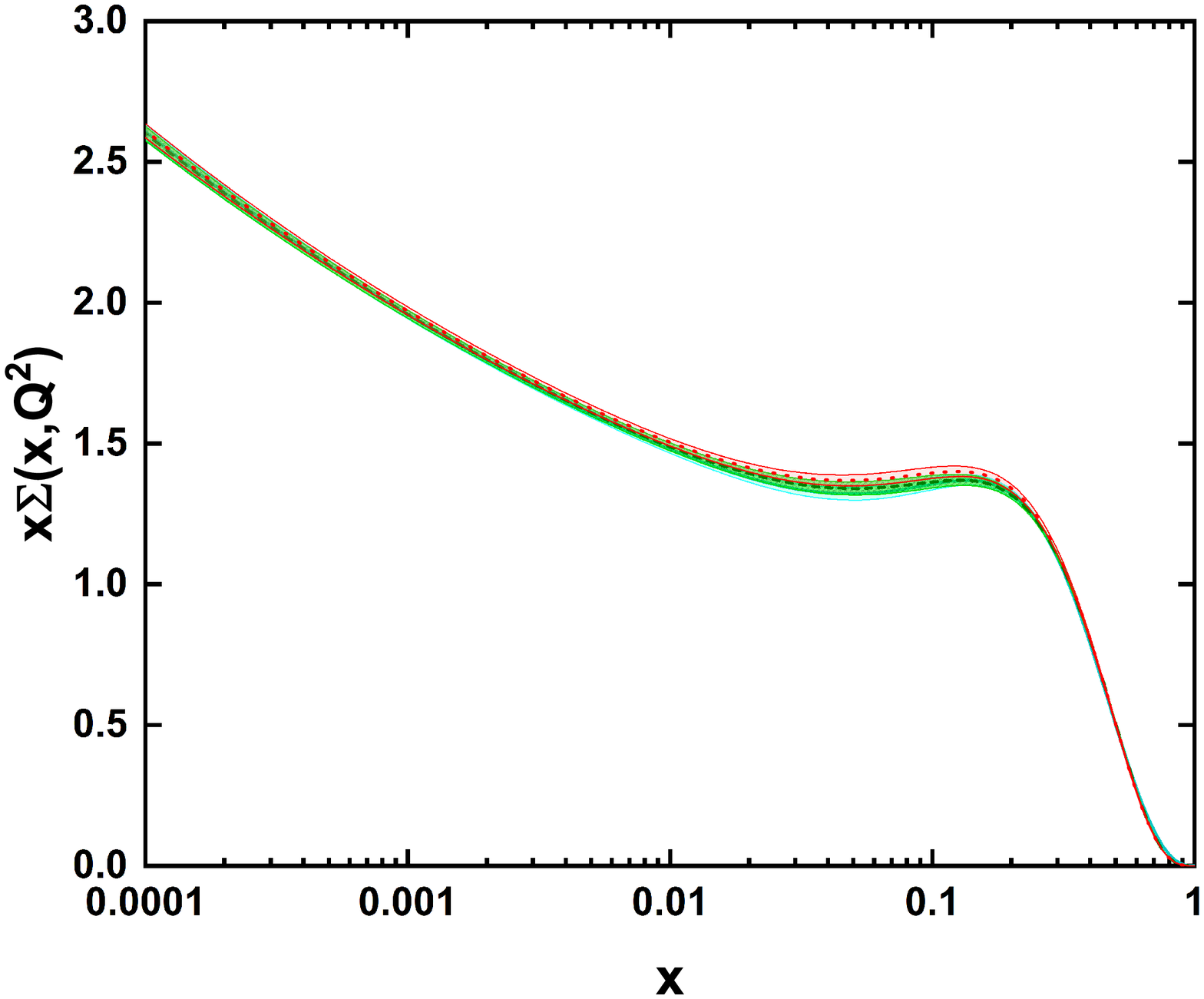}
\includegraphics[width=0.47\textwidth]{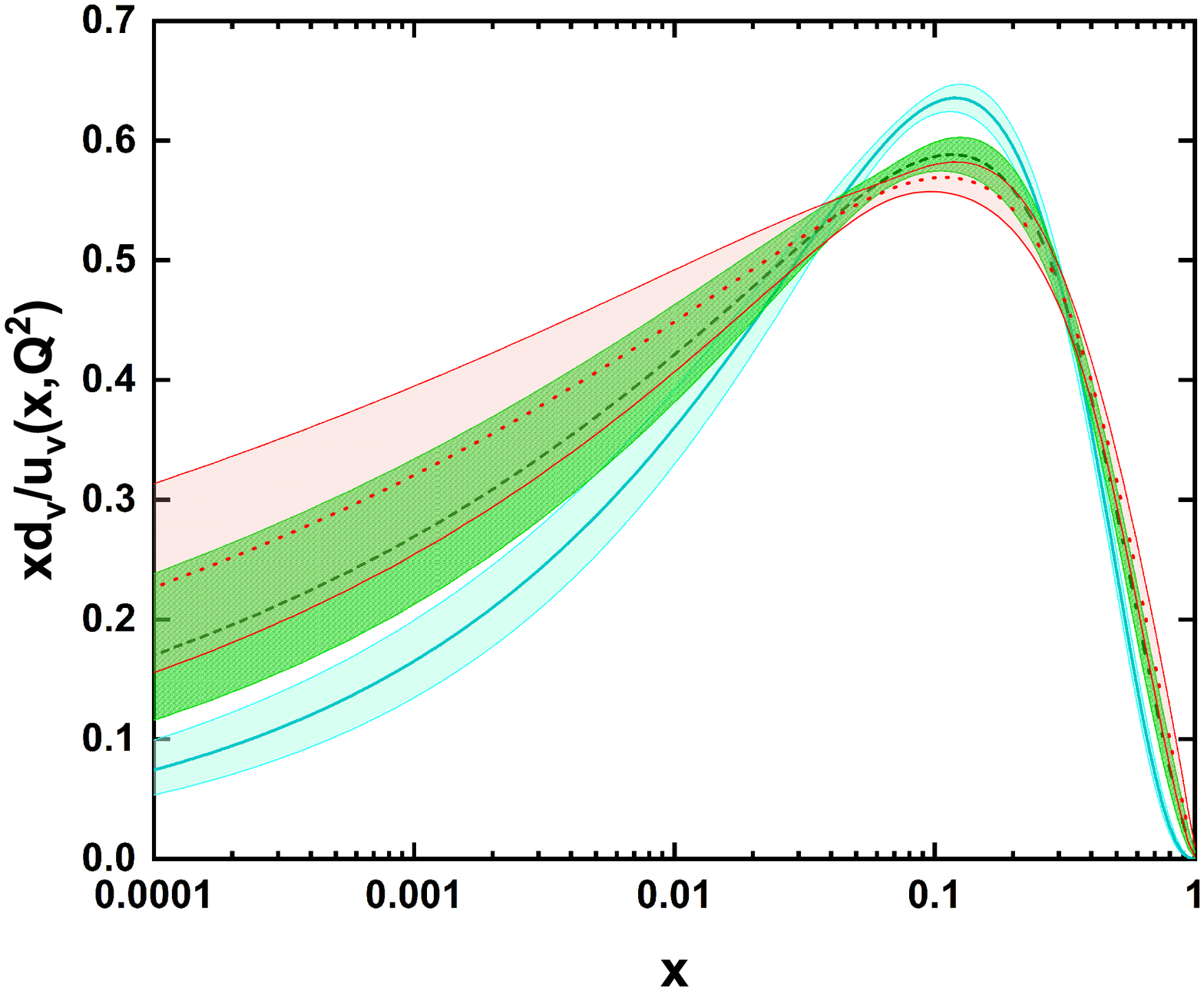}
\caption{A comparison between the $ u_v(x) $ and $ d_v(x) $ valence, gluon $ g(x) $, sum of sea quark $ \Sigma(x) $ and $ d_v(x)/u_v(x) $ distributions at the initial scale $ Q^2_0=1.69 $ GeV$ ^2 $ obtained from three analyses introduced in Sec.~\ref{sec:four}: the \texttt{Base} analysis without considering TMCs and HT (solid curve), \texttt{TMC} analysis including TMCs (dashed curve), and \texttt{TMC\&HT} analysis including TMCs and HT effects simultaneously (dotted curve). The bands represent the related uncertainties.}
\label{fig:fig1}
\end{figure}

\begin{figure}[t!]
\centering
\includegraphics[width=0.47\textwidth]{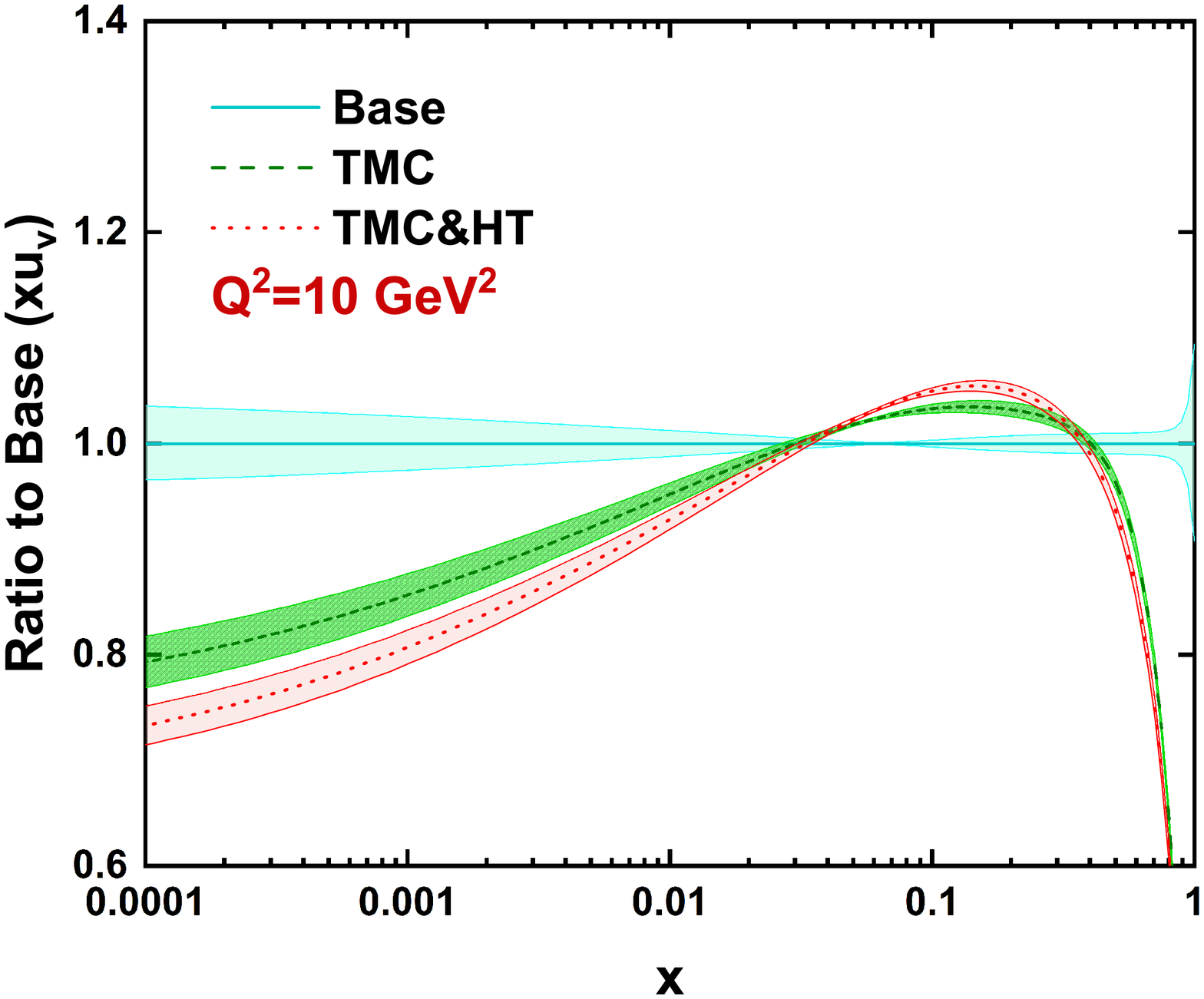}
\includegraphics[width=0.47\textwidth]{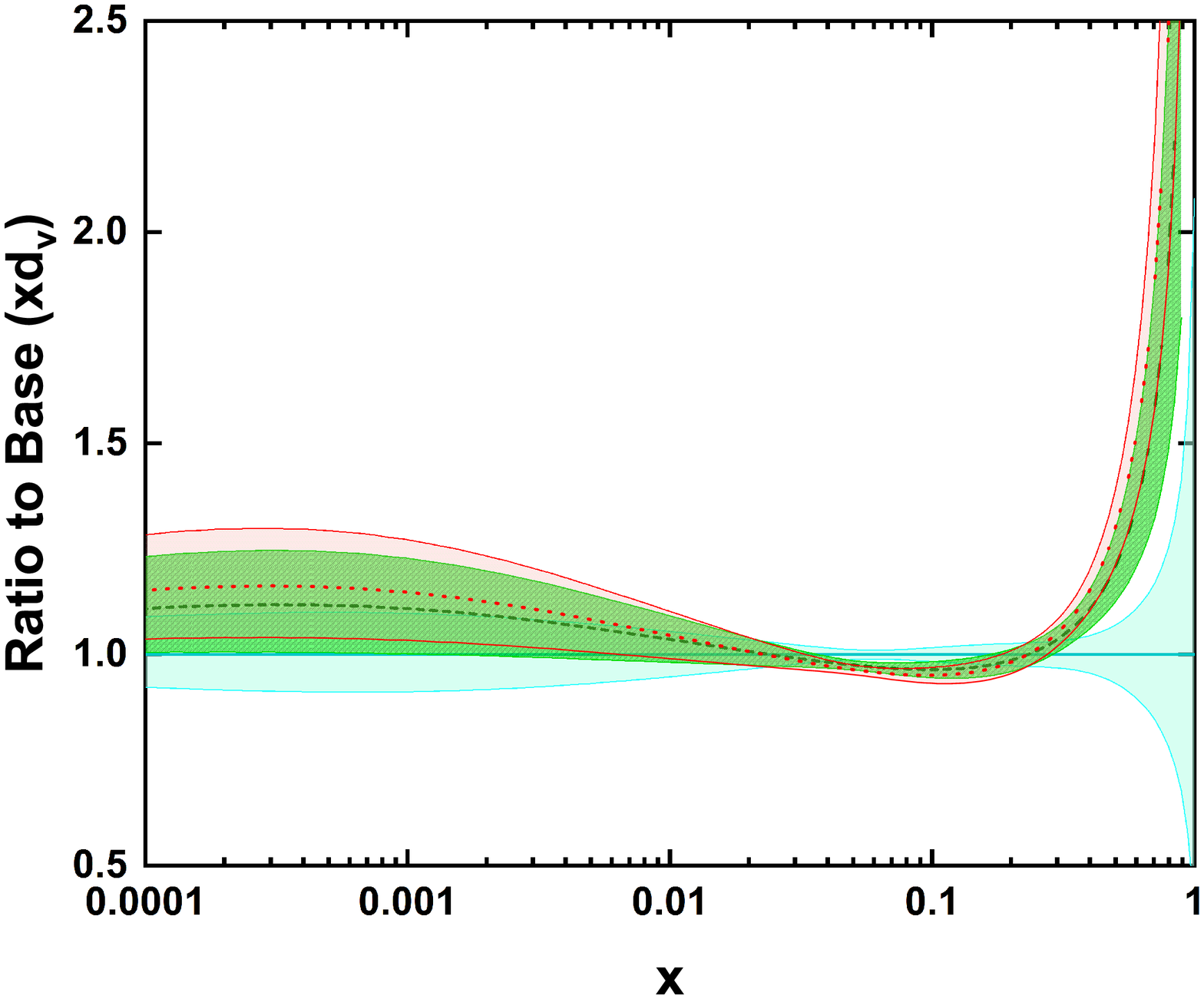}
\includegraphics[width=0.47\textwidth]{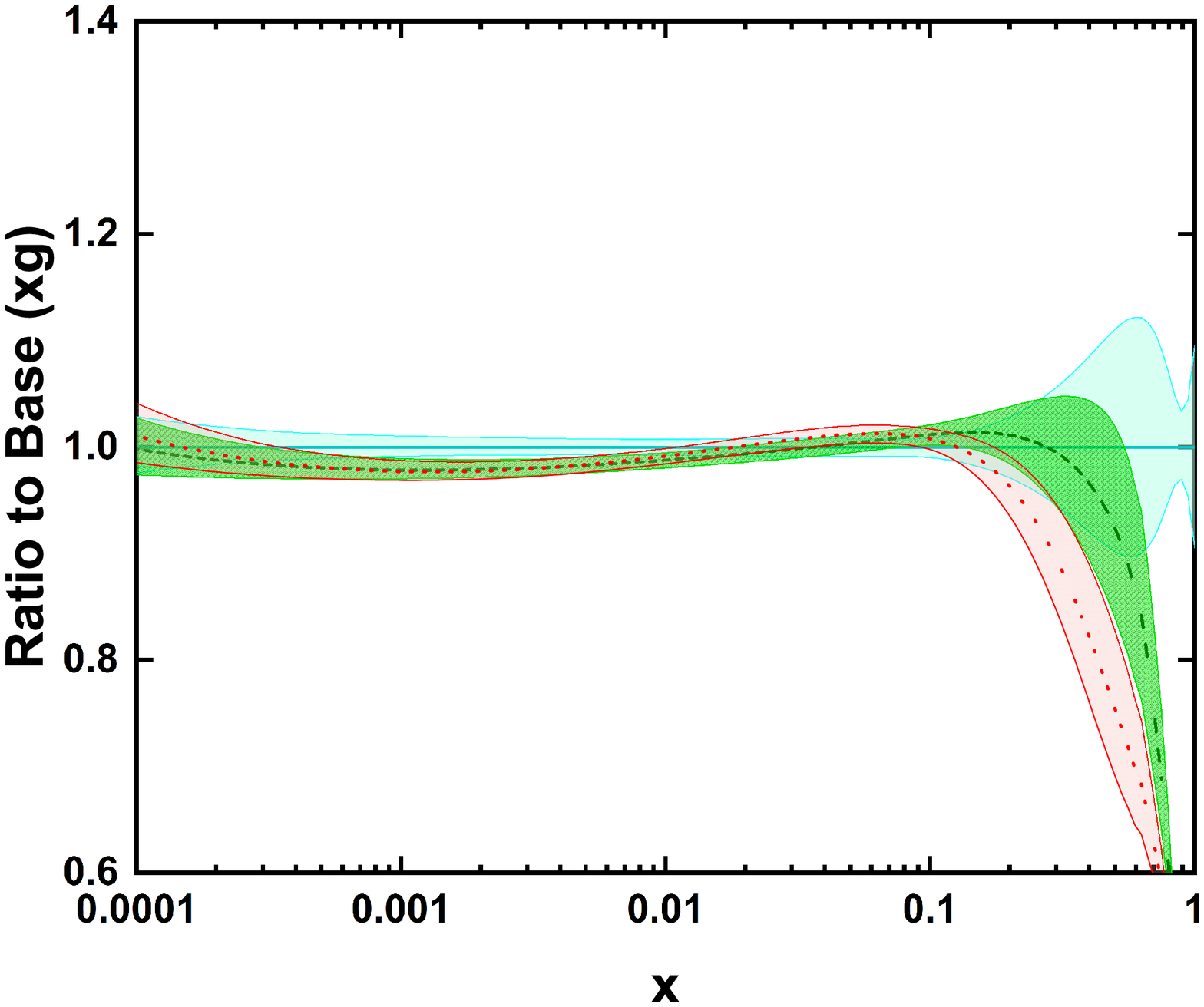}
\includegraphics[width=0.47\textwidth]{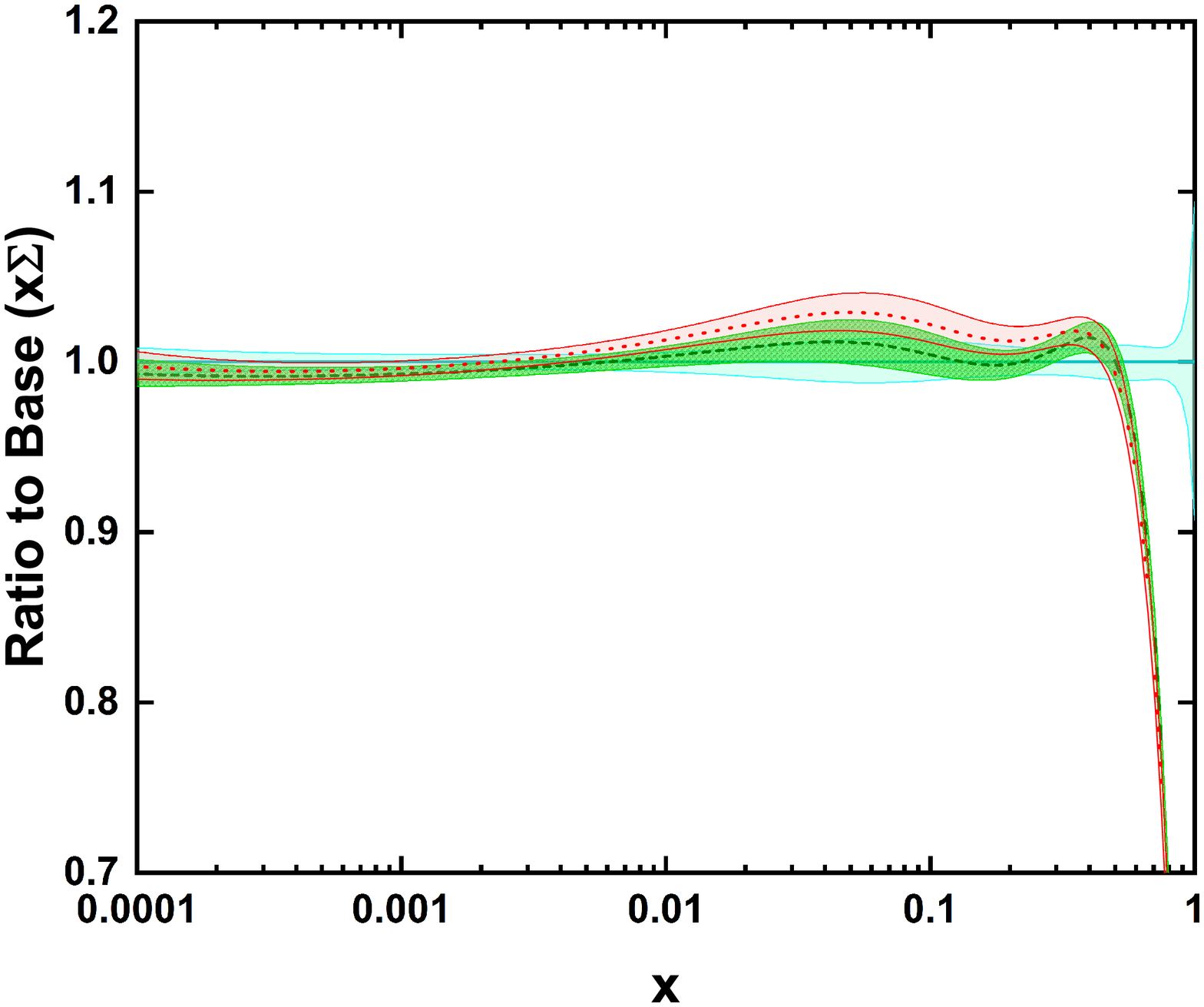}
\includegraphics[width=0.47\textwidth]{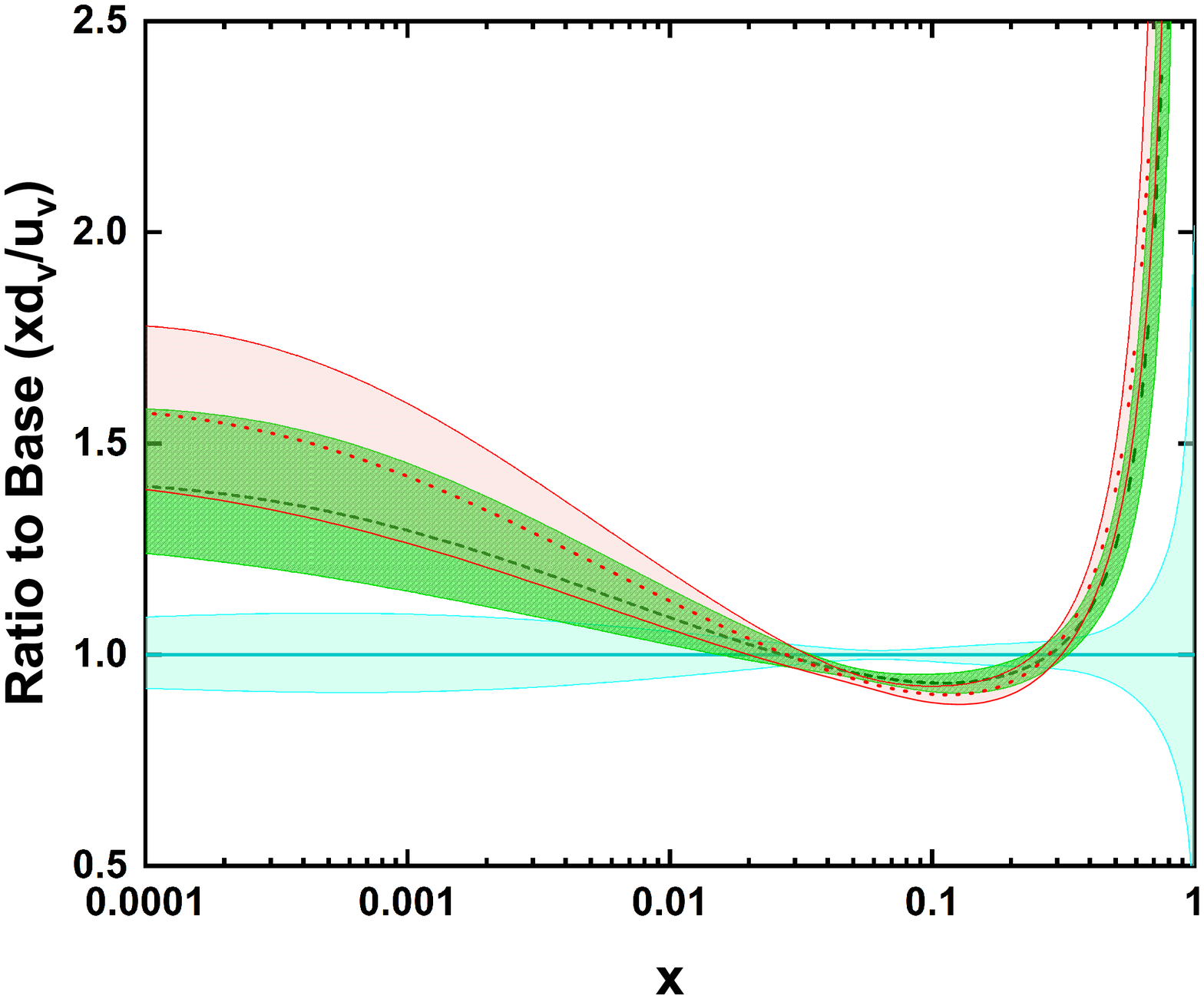}
\caption{A comparison between the $ u_v(x) $ and $ d_v(x) $ valence, gluon $ g(x) $, sum of sea quark $ \Sigma(x) $ and $ d_v(x)/u_v(x) $ distributions at  $ Q^2_0=10 $ GeV$ ^2 $ obtained from three analyses introduced in Sec.~\ref{sec:four}: the \texttt{Base} analysis without considering TMCs and HT (solid curve), \texttt{TMC} analysis including TMCs (dashed curve), and \texttt{TMC\&HT} analysis including TMCs and HT effects simultaneously (dotted curve). The bands represent the related uncertainties. Note that the results have been plotted as ratios to the \texttt{Base} analysis.}
\label{fig:fig2}
\end{figure}

Although the results obtained in the previous section implicitly shows the validity of the calculations, 
in this section we are going to do some cross-checks and scrutinize the results obtained by comparing three analyses \texttt{Base}, \texttt{TMC} and \texttt{TMC\&HT} in different situations. Actually, our motivation comes from the fact that according to the results presented in the previous section (see Figs.~\ref{fig:fig1} and~\ref{fig:fig2}), there are significant changes between the \texttt{Base} and \texttt{TMC} (\texttt{TMC\&HT}) analyses, especially for the up valence distribution, which may be inconsistent at first glance. To understand this point further, it is useful to perform a fit to the same datasets, but applying a more restrictive cut on $ W^2 $ (in line with that commonly applied by global fitters) to remove the data points that are sensitive to the TMCs and HT effects (we do not impose any new cut on $ Q^2 $ and leave it as before as $ Q^2 > 1.69 $ GeV$ ^2 $). In this way, the result of the \texttt{Base} analysis should be more consistent within errors with the \texttt{TMC} (\texttt{TMC\&HT}) case.

\begin{figure}[t!]
\centering
\includegraphics[width=0.47\textwidth]{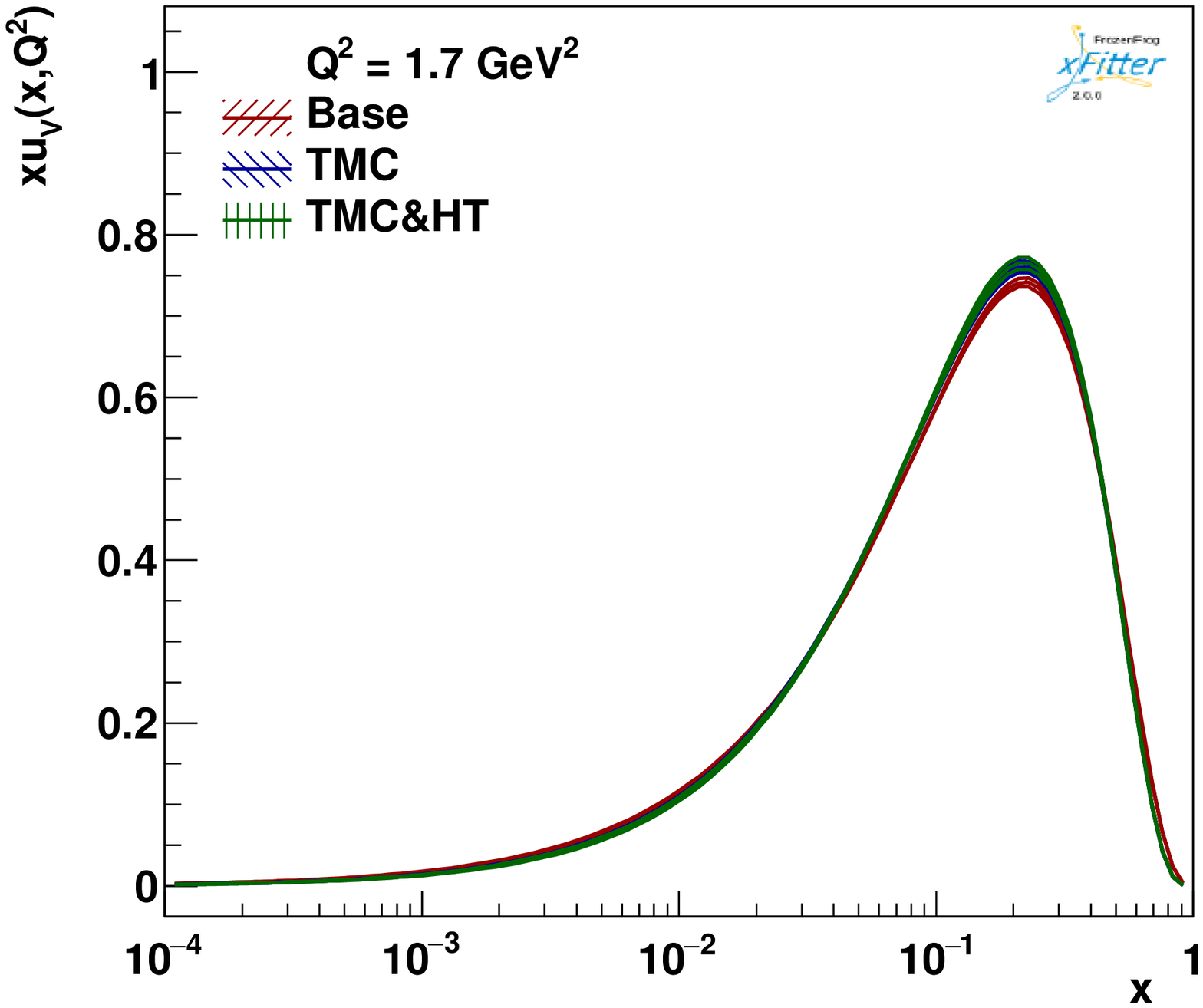}
\includegraphics[width=0.47\textwidth]{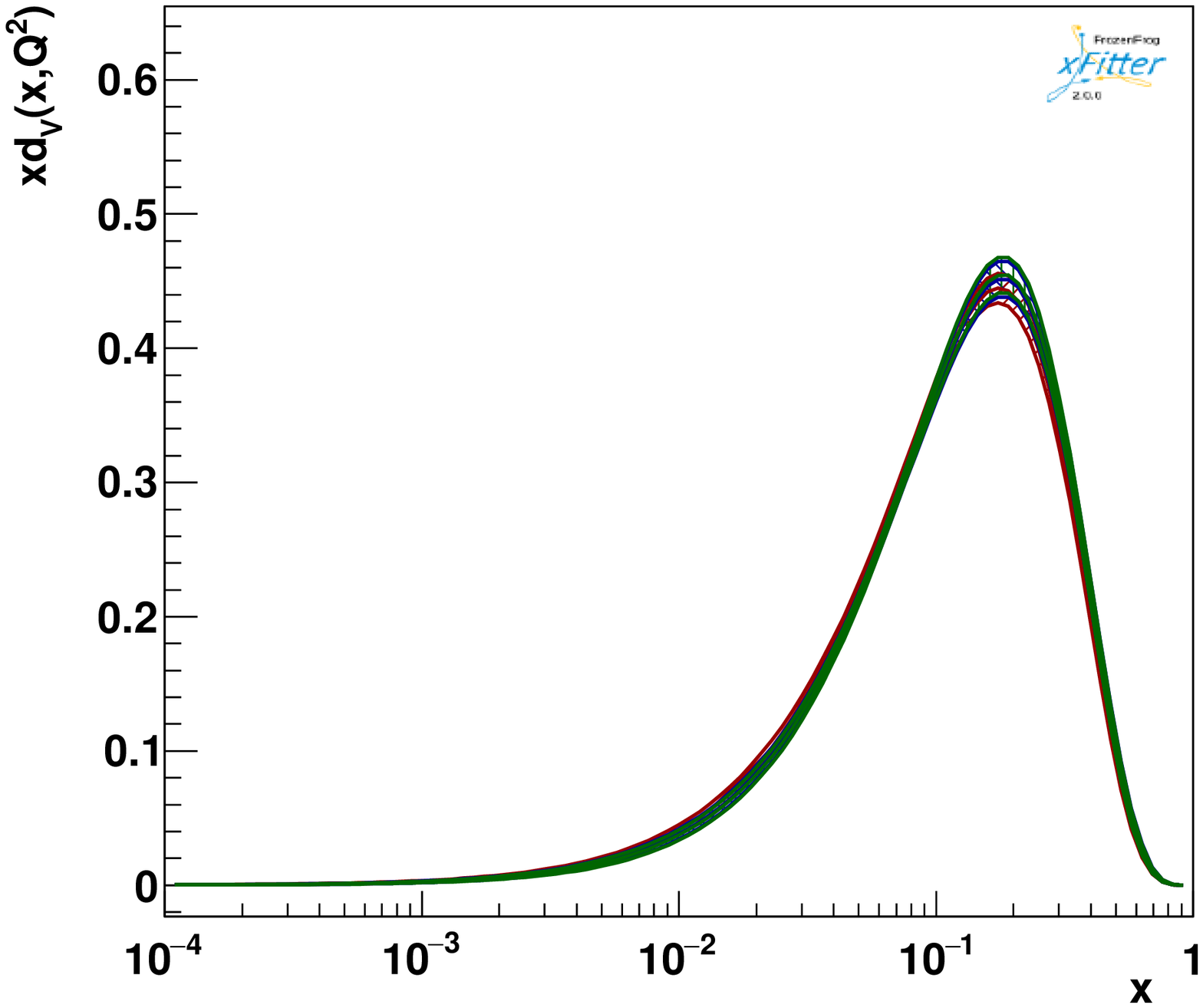}
\includegraphics[width=0.47\textwidth]{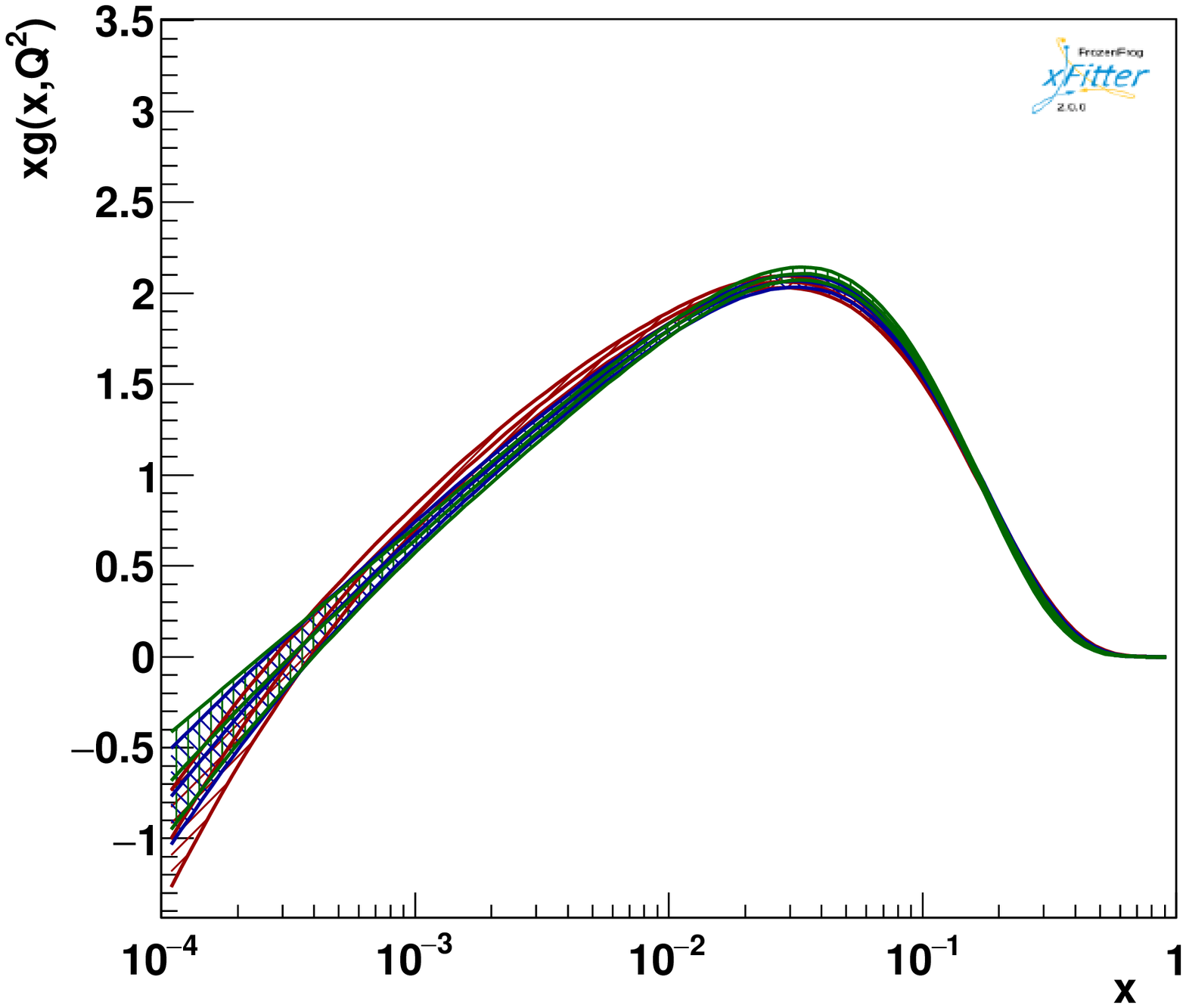}
\includegraphics[width=0.47\textwidth]{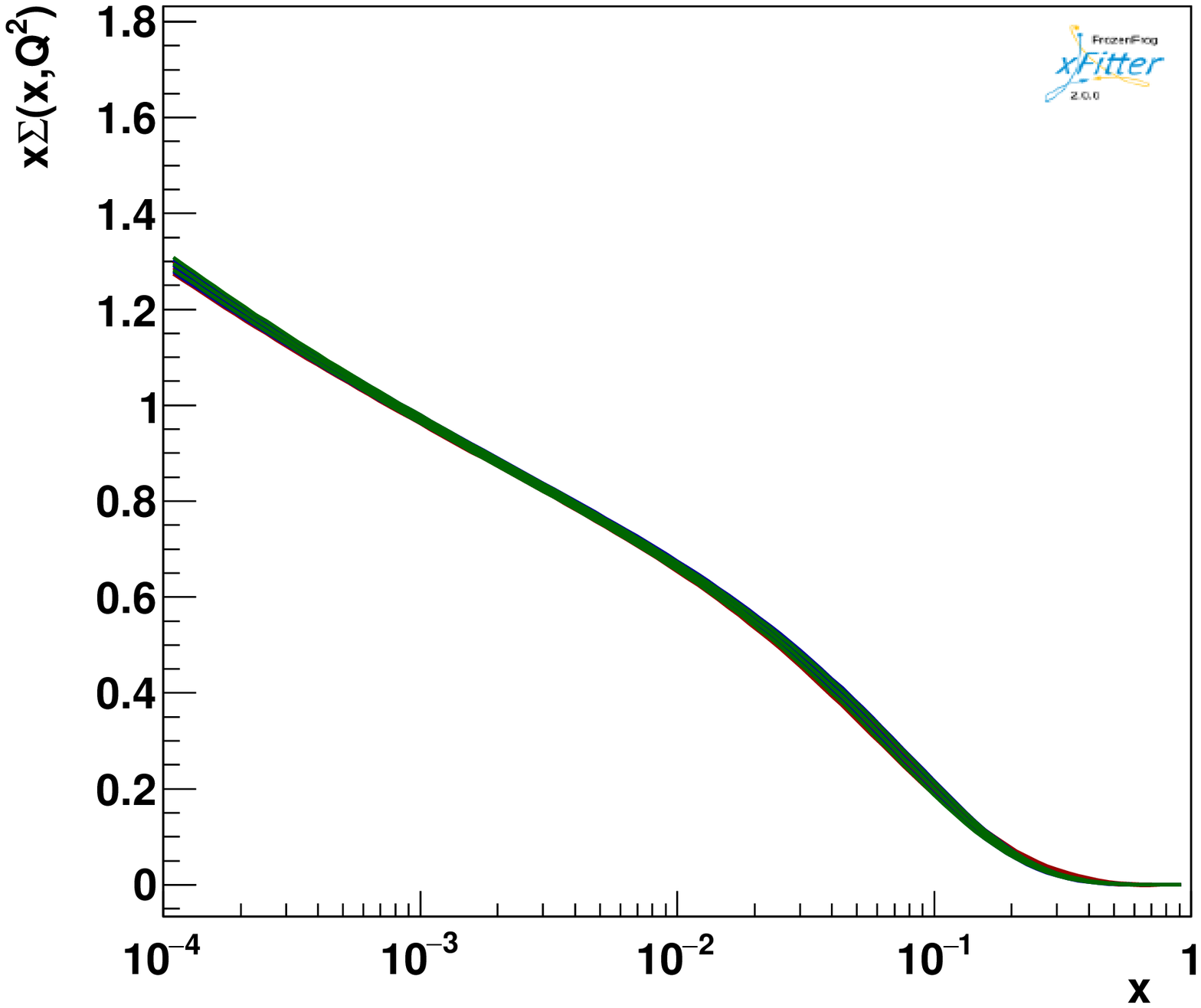}
\includegraphics[width=0.47\textwidth]{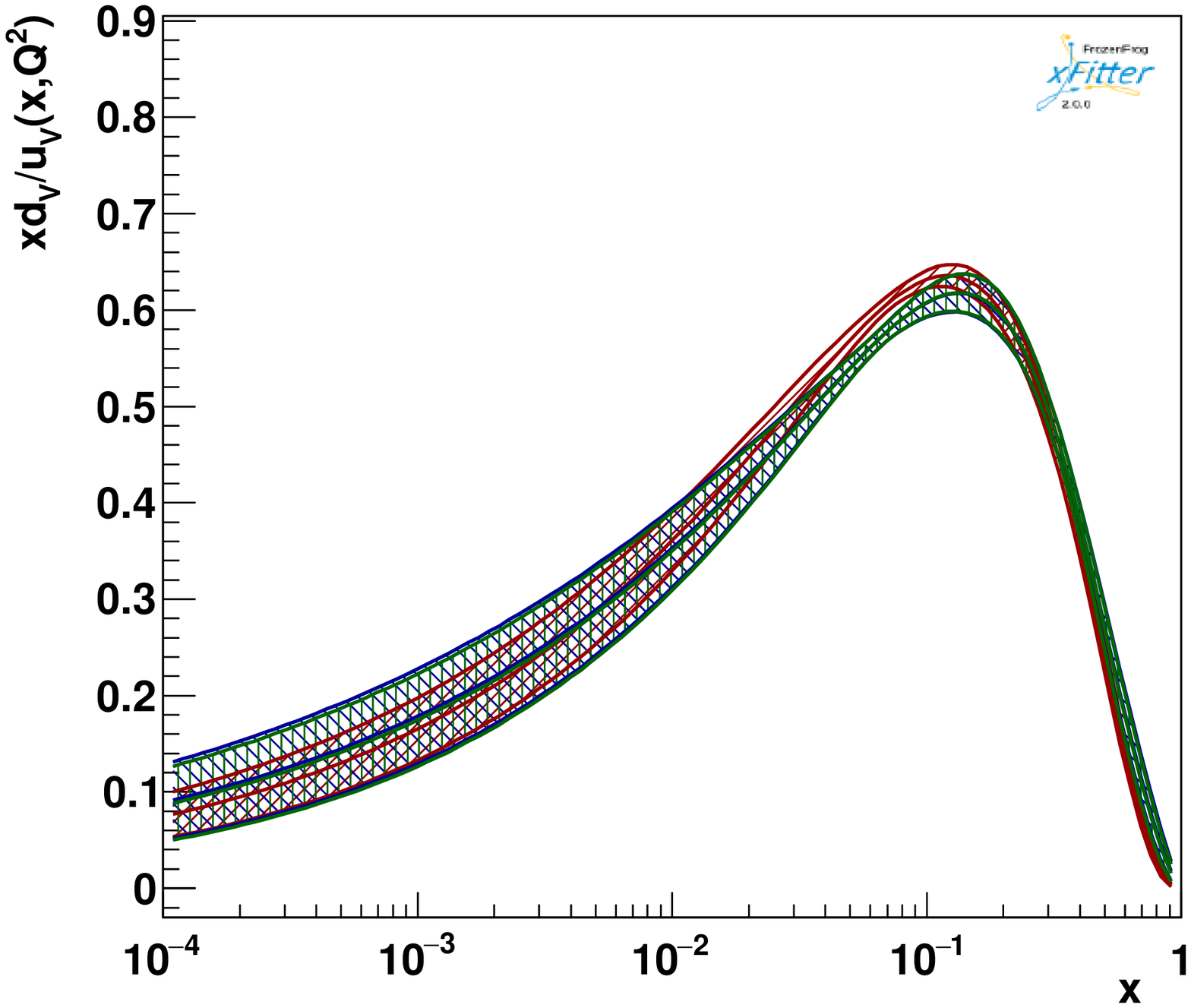}
\caption{A comparison between the $ u_v(x) $ and $ d_v(x) $ valence, gluon $ g(x) $, sum of sea quark $ \Sigma(x) $ and $ d_v(x)/u_v(x) $ distributions at the initial scale $ Q^2_0=1.69 $ GeV$ ^2 $ obtained from  the \texttt{Base} analysis with less restrictive cut $ W^2 > 3$ GeV$ ^2 $ and the \texttt{TMC} and \texttt{TMC\&HT} analyses with more restrictive cut $ W^2 > 15$ GeV$ ^2 $.}
\label{fig:fig3}
\end{figure}

In this regard, we first impose a new cut as $ W^2 > 15$ GeV$ ^2 $ on the data points for the \texttt{TMC} and \texttt{TMC\&HT} analyses and compare the results obtained with the corresponding ones from the previous \texttt{Base} analysis with less restrictive cut $ W^2 > 3$ GeV$ ^2 $. Actually, we expect that by removing the data points that are sensitive to the TMCs and HT effects from the \texttt{TMC} and \texttt{TMC\&HT} analyses their impact on the extracted PDFs is neutralized. Figure~\ref{fig:fig3} shows a comparison between the \texttt{Base} analysis with a less restrictive cut $ W^2 > 3$ GeV$ ^2 $ and the \texttt{TMC} and \texttt{TMC\&HT} analyses with more restrictive cut $ W^2 > 15$ GeV$ ^2 $. As can be seen,   in this case, the result of the \texttt{Base} analysis is more consistent within errors with the \texttt{TMC} and \texttt{TMC\&HT} analyses compared with the previous case (see Fig.~\ref{fig:fig1}). Therefore, one can conclude that the results presented in the previous section that show significant changes between the \texttt{Base} and \texttt{TMC} (\texttt{TMC\&HT}) cases are admissible.

\begin{figure}[t!]
\centering
\includegraphics[width=0.47\textwidth]{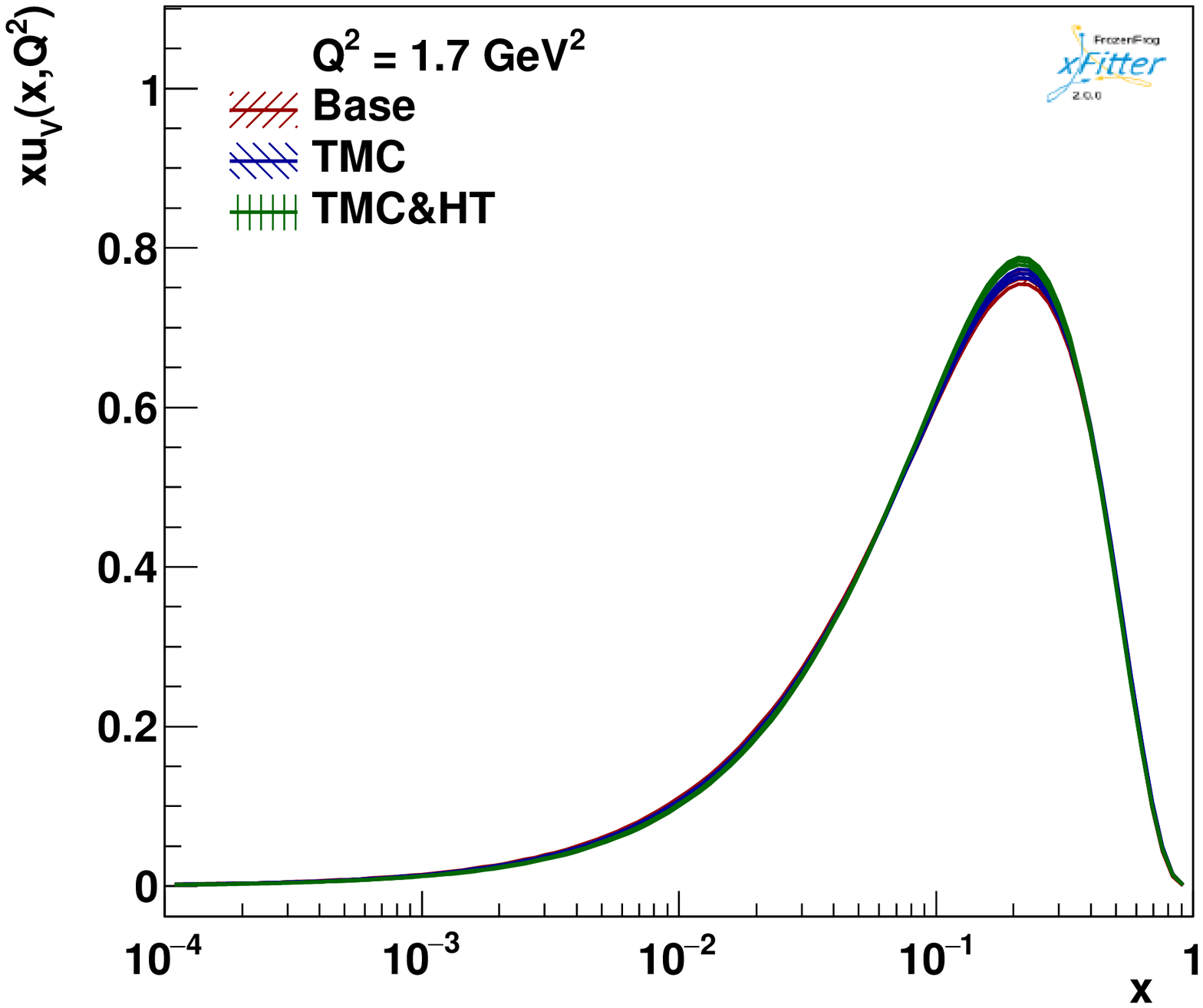}
\includegraphics[width=0.47\textwidth]{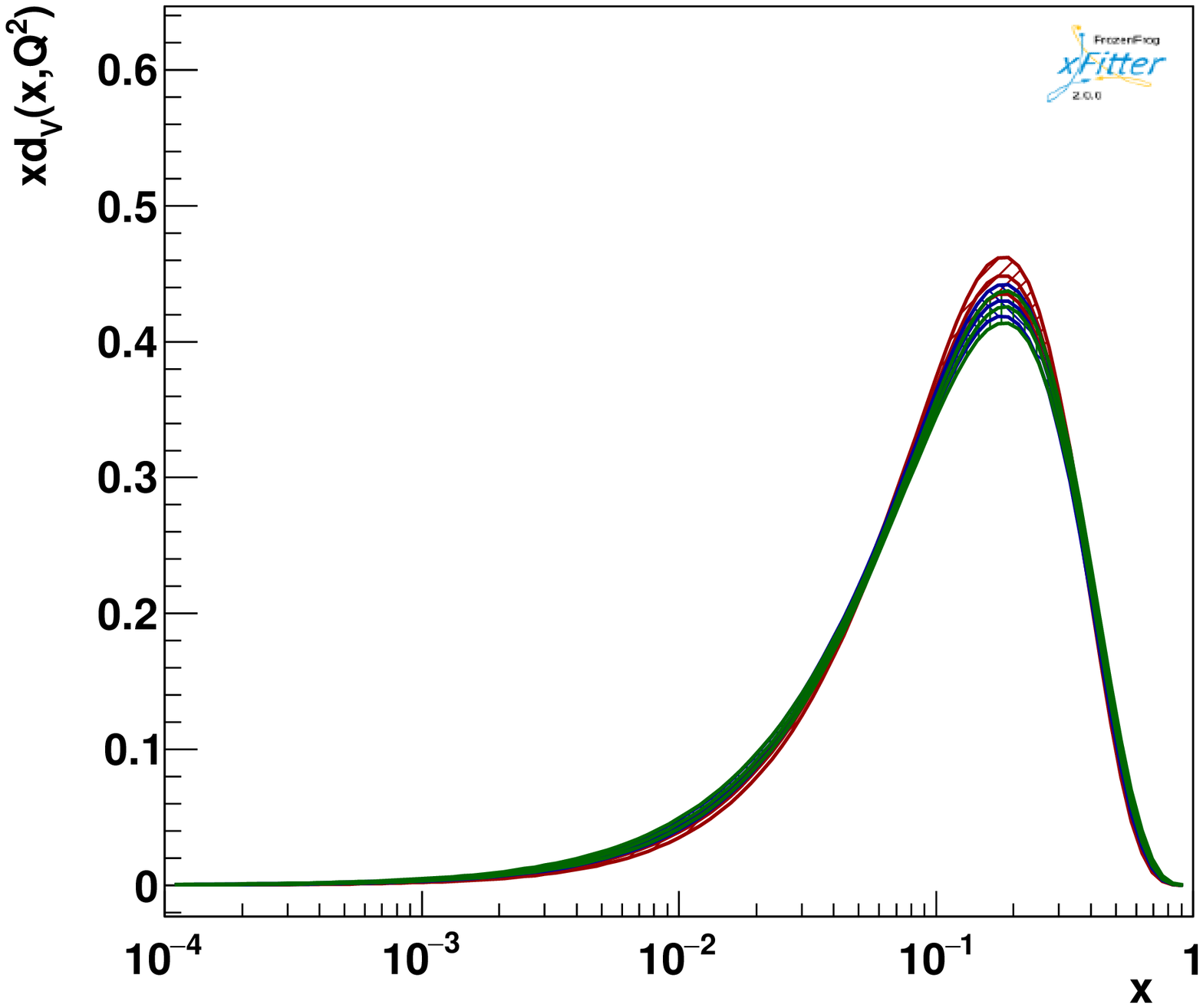}
\includegraphics[width=0.47\textwidth]{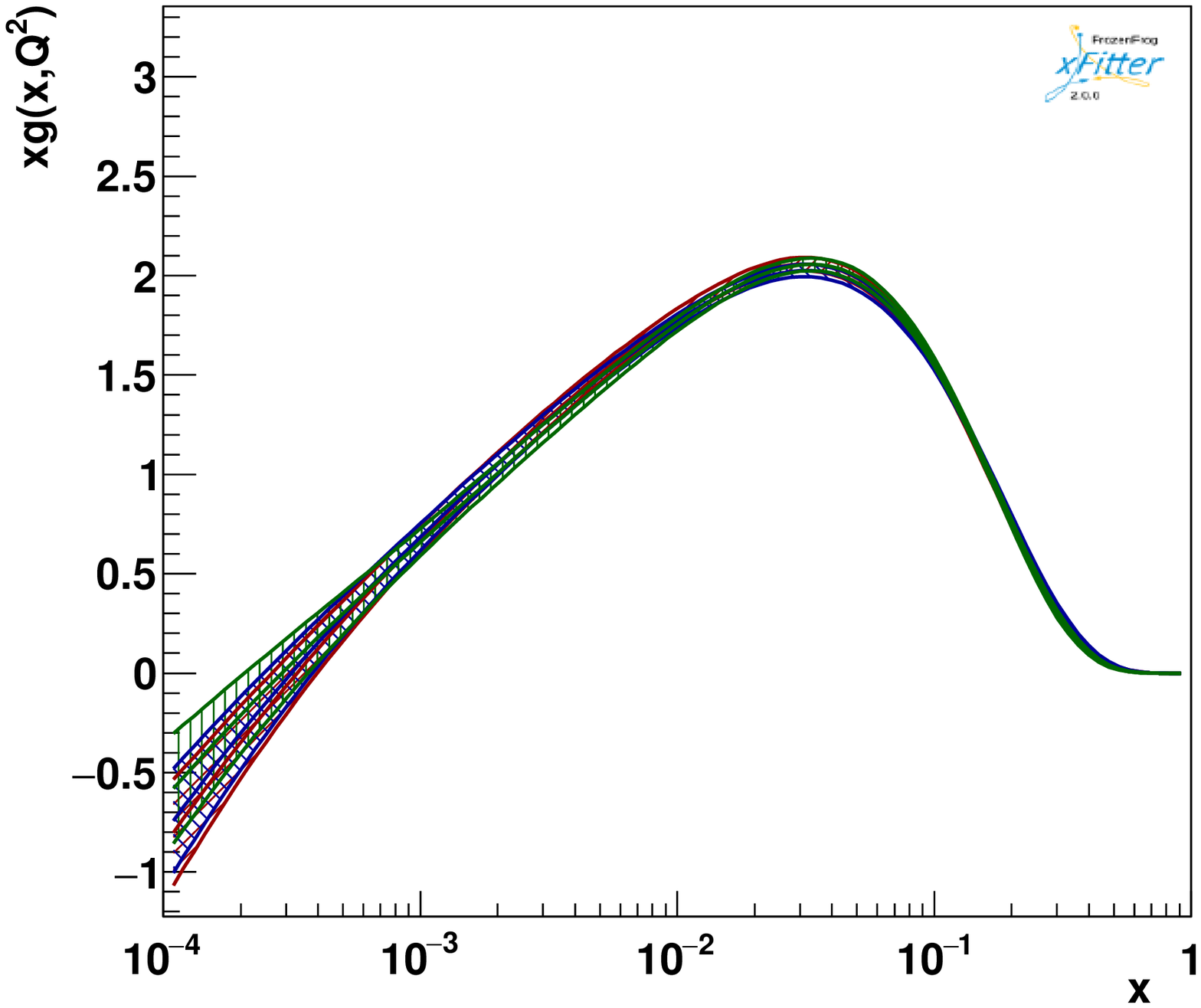}
\includegraphics[width=0.47\textwidth]{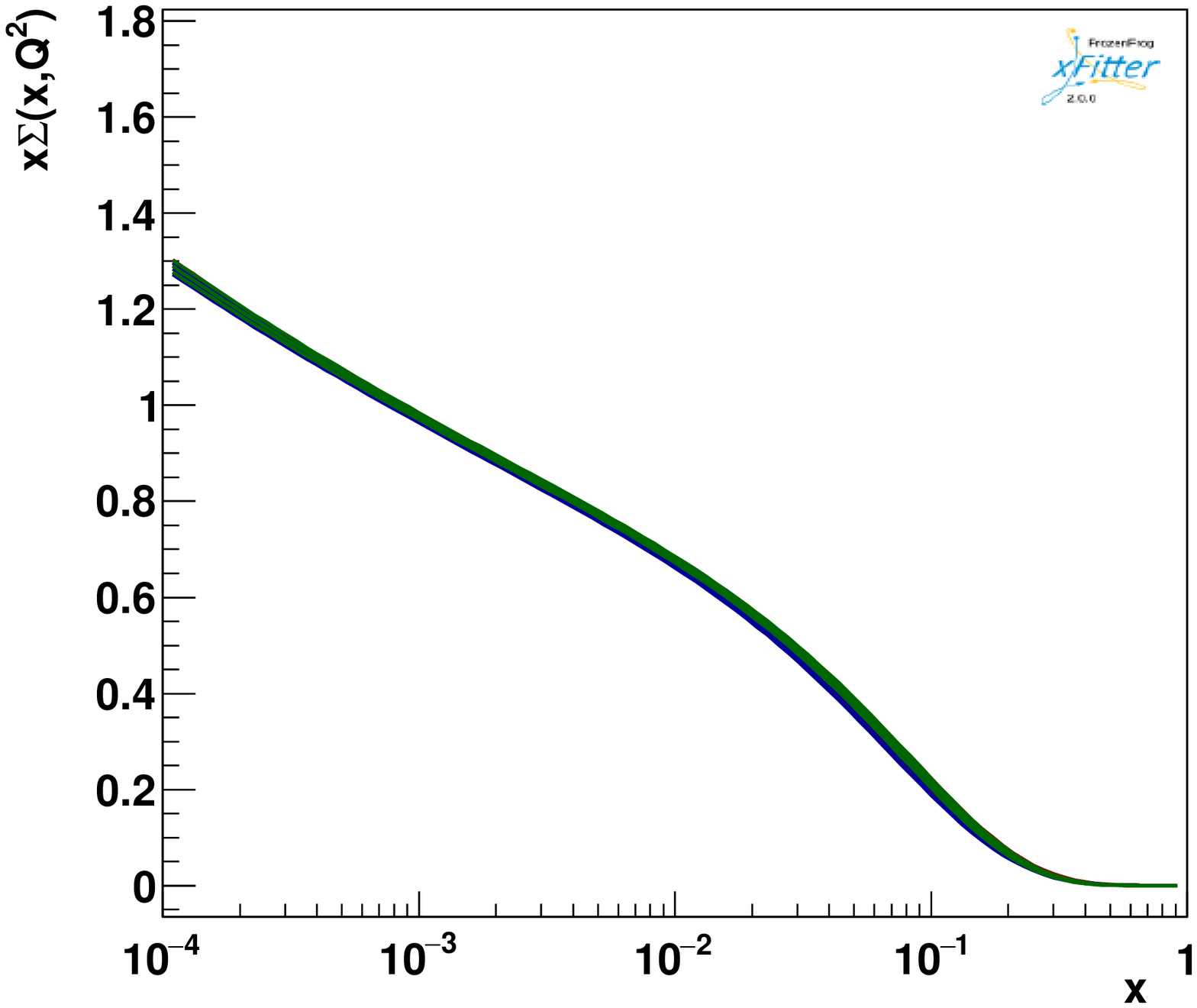}
\includegraphics[width=0.47\textwidth]{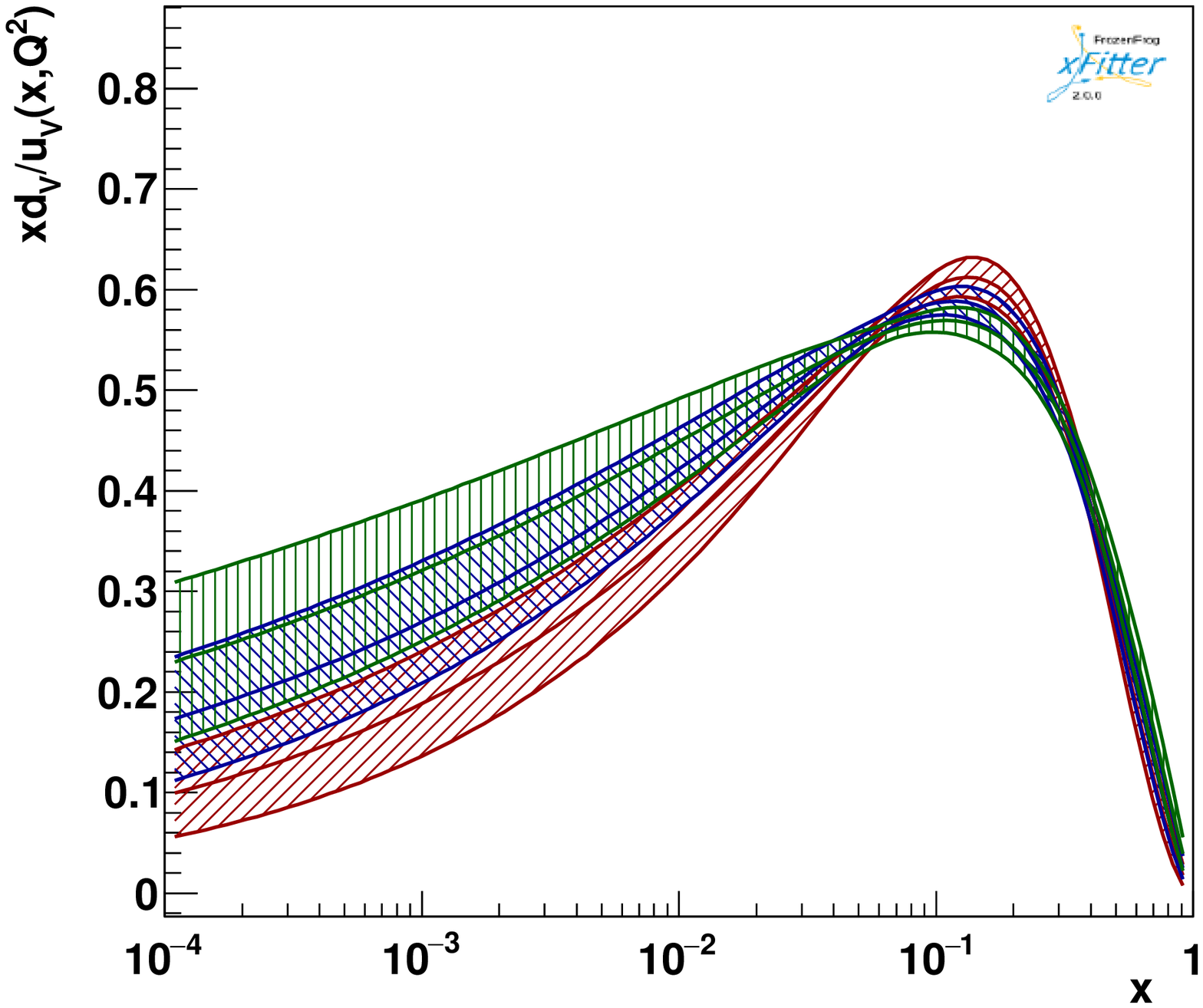}
\caption{A comparison between the $ u_v(x) $ and $ d_v(x) $ valence, gluon $ g(x) $, sum of sea quark $ \Sigma(x) $ and $ d_v(x)/u_v(x) $ distributions at the initial scale $ Q^2_0=1.69 $ GeV$ ^2 $ obtained from  the \texttt{Base} analysis with more restrictive cut $ W^2 > 15$ GeV$ ^2 $ and the \texttt{TMC} and \texttt{TMC\&HT} analyses with less restrictive cut $ W^2 > 3$ GeV$ ^2 $.}
\label{fig:fig4}
\end{figure}

As another study, this time we perform a new \texttt{Base} analysis with a more restrictive cut $ W^2 > 15$ GeV$ ^2 $ imposed on data points and compare its results with the corresponding ones from the \texttt{TMC} and \texttt{TMC\&HT} analyses with a less restrictive cut $ W^2 > 3$ GeV$ ^2 $ obtained in the previous section. Again we expect that the result of the \texttt{Base} analysis will be more consistent within errors with the \texttt{TMC} and \texttt{TMC\&HT} analyses, but with larger errors. Figure~\ref{fig:fig4} shows a comparison between the \texttt{Base} analysis with a more restrictive cut $ W^2 > 15$ GeV$ ^2 $ and the \texttt{TMC} and \texttt{TMC\&HT} analyses with a less restrictive cut $ W^2 > 3$ GeV$ ^2 $. If one compares the up valence distribution from Figs.~\ref{fig:fig1} and~\ref{fig:fig4}, it becomes clear that by removing the data points with $ W^2 < 15$ GeV$ ^2 $ in the \texttt{Base} analysis, the TMCs and HT effects have no previous impact on the extracted PDFs and the results of three analyses become more consistent. Therefore, the observed changes in PDFs from the \texttt{Base} analysis to the \texttt{TMC} (\texttt{TMC\&HT}) one in Figs.~\ref{fig:fig1} and~\ref{fig:fig2} are really due to considering the TMCs and HT effects in the analysis.

%
\section{Summary and conclusions}\label{sec:six}

In any global analysis of PDFs, target mass corrections (TMCs) and higher-twist (HT) effects are most important modifications on the DIS structure functions which should be considered when the analysis includes also data points with larger values of $ x $ and smaller values of photon virtuality $ Q^2 $, where nonperturbative QCD effects generally play a more significant role. In this work, we reported and discussed our recent activities on the implementation of TMCs and HT effects in the xFitter framework, as a comprehensive, powerful and open-source tool for studying PDFs. All developments have been performed in a user-friendly way, so that the user can switch on or off TMCs and HT, and also choose between different options. 

For the case of TMCs, we have added three options to the xFitter package: ``\texttt{Approximate}," `\texttt{Exact}" and ``\texttt{APFEL.}" The \texttt{APFEL} TMCs just works with the \texttt{FONLL} heavy quark scheme, while the user can choose any heavy flavor scheme if the \texttt{Exact} or \texttt{Approximate} TMC is selected. However, \texttt{APFEL} has an advantage to perform related calculations up to 3 times faster than \texttt{Exact} TMCs. As we checked it, using \texttt{APFEL} and \texttt{Exact} options leads to the results which are in an excellent agreement with each other, though \texttt{Exact} leads to a somewhat better fit of DIS data than \texttt{APFEL}. For the case of HT, we have added the ``\texttt{CJ15}" option to the xFitter package which parametrizes HT effects as a function of $ x $ and uses a commonly used phenomenological form for the total structure function. 

We checked the validity of the modifications by performing sample analyses of PDFs with and without considering TMCs and HT effects and using a wide range of DIS data.
We affirm that considering such corrections can significantly change the extracted PDFs, especially the valence distributions, and improve the agreement between the theory and experiment. To be more precise, 
the $ u_v(x) $ and $ d_v(x) $ valence distributions are more affected by including TMCs and HT, since they are dominant at larger values of $ x $ than the others where TMCs and HT effects are also effective.
The developed xFitter package including TMCs and HT can be useful for further studies in this subject. It can be found on the \texttt{Gitlab} repository~\cite{gitlab}.

%
\acknowledgments

We thank the xFitter developers, especially Alexander Glazov, Fred Olness, Oleksandr Zenaiev and Valerio Bertone, for helpful discussions. Muhammad Goharipour thanks also the School of Particles and Accelerators, Institute for Research in Fundamental Sciences (IPM) for financial support provided for this research.
%

%

\newpage

\begin{table}[th!]
\caption{The results of three analyses introduced at the beginning of Sec.~\ref{sec:four}: the \texttt{Base} analysis without considering TMCs and HT, \texttt{TMC} analysis including TMCs, and \texttt{TMC\&HT} analysis including TMCs and HT effects simultaneously which have been presented in the third, fourth and fifth columns, respectively. The analyses have been performed at NNLO using the \texttt{FONLL-C} heavy quark scheme.}\label{tab:chi2}
\centering
{\scriptsize 
\newcolumntype{C}[1]{>{\hsize=#1\centering\arraybackslash}X}
\centering
\begin{tabularx}{0.9\linewidth}{llc*{4}{C{3cm}C{3cm}}C{3cm}}  \hline \hline
Observable
  & Experiment &  \multicolumn{3}{c}{ $\chi^2$/$ N_{\textrm{pts.}} $  } \\
  &  		& \texttt{Base}      & \texttt{TMC}  & \texttt{TMC\&HT} \\
 \hline
 
DIS $\sigma$	
 &   HERA1+2 NCep 820 \cite{Abramowicz:2015mha}& 87 / 75& 89 / 75& 90 / 75   \\ 
 & HERA1+2 NCep 920 \cite{Abramowicz:2015mha}& 562 / 402& 535 / 402& 536 / 402   \\ 
 & HERA1+2 NCep 460 \cite{Abramowicz:2015mha}& 234 / 209& 232 / 209& 231 / 209   \\ 
&  HERA1+2 NCep 575\cite{Abramowicz:2015mha} &  236 / 259& 237 / 259& 237 / 259   \\ 
 & HERA1+2 CCep \cite{Abramowicz:2015mha}&54 / 39& 52 / 39& 53 / 39 \\ 
  &HERA1+2 CCem \cite{Abramowicz:2015mha}&52 / 42& 64 / 42& 71 / 42  \\ 
  &HERA1+2 NCem \cite{Abramowicz:2015mha} &  272 / 159& 240 / 159& 237 / 159  \\ 
 & HERA c \cite{Abramowicz:1900rp}& 61 / 52& 65 / 52& 65 / 52 \\ 
 & HERA b \cite{Abramowicz:2014zub}&  20 / 27& 21 / 27& 21 / 27   \\ 
  DIS $F_2$
  &BCDMS 100~GeV \cite{Benvenuti:1989rh}& 129 / 97& 112 / 97& 109 / 97 \\ 
    &BCDMS 120~GeV \cite{Benvenuti:1989rh}   &109 / 99& 83 / 99& 83 / 99 \\ 
 &  BCDMS  200~GeV \cite{Benvenuti:1989rh}&  166 / 79& 94 / 79& 95 / 79  \\ 
&  BCDMS  280~GeV \cite{Benvenuti:1989rh}&81 / 75& 78 / 75& 78 / 75  \\ 
&SLAC \cite{Whitlow:1991uw}& 1353 / 565& 674 / 565& 622 / 565  \\ 
&NMC 90~GeV \cite{Arneodo:1996qe}& 72 / 62& 69 / 62& 73 / 62  \\ 
&NMC 120~GeV \cite{Arneodo:1996qe}& 104 / 63& 105 / 63& 105 / 63   \\ 
&NMC 200~GeV\cite{Arneodo:1996qe} & 108 / 72& 106 / 72& 102 / 72 \\ 
&NMC 280~GeV \cite{Arneodo:1996qe}&151 / 78& 151 / 78& 152 / 78  \\ 
&E665 \cite{Adams:1996gu}& 57 / 53& 57 / 53& 56 / 53  \\ 
&JLab \cite{Malace:2009kw}&  583 / 142& 317 / 142& 292 / 142 \\ \hline
    Correlated $\chi^2$  && 325& 247& 224    \\ 
  Log penalty $\chi^2$  &&   +91& +27& +16  \\ 
  \hline
  Total $\chi^2$/d.o.f.  && 4908 / 2635& 3654 / 2635& 3548 / 2632
   \\ \hline \hline\\ 
\end{tabularx}
}
\end{table}

\newpage
%
\begin{table}[th!]
\caption{The optimum parameters obtained from different analyses introduced in Sec.~\ref{sec:four} at initial scale $ Q_0=1.3 $ GeV: the \texttt{Base} analysis without considering TMCs and HT, \texttt{TMC} analysis including TMCs, and \texttt{TMC\&HT} analysis including TMCs and HT effects simultaneously which have been presented in the second, third and fourth columns, respectively. The parameters in bold type have been fixed in the fit.}\label{tab:par}
\centering
{\scriptsize 
\newcolumntype{C}[1]{>{\hsize=#1\centering\arraybackslash}X}
\centering
\begin{tabularx}{0.8\linewidth}{c*{3}{C{3.6cm}C{3.6cm}}C{3.6cm}}  \hline \hline

\hline
 Parameter   & \texttt{Base} & \texttt{TMC} & \texttt{TMC\&HT}  \\ 
\hline

\hline
  $ B_g $ &  $0.028 \pm 0.039$& $0.120 \pm 0.036$& $0.198 \pm 0.036$   \\ 
  $ C_g $ & $6.27 \pm 0.37$& $6.64 \pm 0.36$& $7.45 \pm 0.40$   \\ 
  $ A^\prime _g $ & $0.296 \pm 0.078$& $0.163 \pm 0.054$& $0.084 \pm 0.037$   \\ 
$ B^\prime_g $ & $-0.271 \pm 0.023$& $-0.282 \pm 0.033$& $-0.315 \pm 0.045$ \\ 
$ C^\prime_g $ & $\bold{ 25.00 }$& $\bold{ 25.00 }$& $\bold{ 25.00 }$   \\ 
 $ B_{u_v} $ & $0.8355 \pm 0.0085$& $0.9117 \pm 0.0086$& $0.9360 \pm 0.0073$ \\ 
  $  C_{u_v} $ & $3.059 \pm 0.038$& $3.833 \pm 0.028$& $3.763 \pm 0.038$  \\ 
 $  E_{u_v} $ & $0.27 \pm 0.11$& $1.73 \pm 0.18$& $1.05 \pm 0.17$\\ 
$  B_{d_v} $& $1.184 \pm 0.050$& $1.111 \pm 0.052$& $1.086 \pm 0.051$  \\ 
$  C_{d_v} $ &$5.55 \pm 0.21$& $5.00 \pm 0.24$& $4.79 \pm 0.23$  \\ 
 $  C_{\bar{U}} $ & $3.66 \pm 0.41$& $6.7 \pm 1.0$& $6.96 \pm 0.94$  \\ 
$  D_{\bar{U}} $ & $-1.55 \pm 0.35$& $-0.61 \pm 0.93$& $-0.84 \pm 0.88$  \\ 
 $  A_{\bar{D}}$ & $0.2598 \pm 0.0059$& $0.2631 \pm 0.0054$& $0.2658 \pm 0.0053$   \\ 
$  B_{\bar{D}} $&  $-0.1241 \pm 0.0030$& $-0.1225 \pm 0.0027$& $-0.1218 \pm 0.0027$   \\ 
$  C_{\bar{D}} $ & $14.1 \pm 1.5$& $11.7 \pm 1.3$& $10.18 \pm 0.98$  \\ 
 $  \alpha_s $& $\bold{ 0.1180 }$& $\bold{ 0.1180 }$& $\bold{ 0.1180 }$  \\ 
$  f_s $& $\bold{ 0.4000 }$& $\bold{ 0.4000 }$& $\bold{ 0.4000 }$  \\ 
  $ h_0 $ & ... & ... & $-1.32 \pm 0.20$  \\ 
  $ h_1 $ & ... & ... & $1.49 \pm 0.16$   \\ 
  $ h_2 $ & ... & ... &$-1.963 \pm 0.071$ \\ 
\hline \hline
\end{tabularx}
}
\end{table}
\newpage

%

\newpage

%

\newpage
%

\newpage
%

\end{document}